\documentclass[twocolumn,twocolappendix]{aastex7}

\usepackage{amsmath}
\usepackage{amssymb}
\usepackage{bm}
\usepackage{enumitem}
\newcommand{\hst}{{\it HST}}
\newcommand{\jwst}{{\it JWST}}

\newcommand{\beagle}{\textsc{Beagle}}

\newcommand{\eazy}{\textsc{Eazy}}

\newcommand{\muv}{$M_{\rm UV}$}

\newcommand{\gtrsm}{\gtrsim}

\newcommand{\lya}{\hbox{Ly$\alpha$}}
\newcommand{\ha}{\hbox{H$\alpha$}}
\newcommand{\hb}{\hbox{H$\beta$}}
\newcommand{\hg}{\hbox{H$\gamma$}}
\newcommand{\oiii}{\hbox{O\,{\sc iii}}}

\newcommand{\oiiihb}{[\oiii{}]+\hb{}}

\newcommand{\ergscmA}{\,erg~s$^{-1}$~cm$^{-2}$~\AA$^{-1}$}

\newcommand{\um}{\,$\mu$m}
\newcommand{\kms}{\,km~s$^{-1}$}

\graphicspath{{./}{figures/}}

\begin{document}

\title{The Impact of Galaxy Overdensities and Ionized Bubbles on \lya{}  Emission at $z\sim7.0\mbox{--}8.5$ }

\shorttitle{\lya{} Emission and Overdensities at $z\sim7.0\mbox{--}8.5$}
\shortauthors{Chen et al.}

\correspondingauthor{Zuyi Chen}
\email{zychen@arizona.edu}

\author[0000-0002-2178-5471]{Zuyi Chen}
\affiliation{Steward Observatory, University of Arizona, 933 N Cherry Ave, Tucson, AZ 85721, USA}
\email{zychen@arizona.edu}  

\author[0000-0001-6106-5172]{Daniel P. Stark}
\affiliation{Department of Astronomy, University of California, 501 Campbell Hall \#3411, Berkeley, CA 94720, USA}
\email{dpstark@berkeley.edu}  

\author[0000-0002-3407-1785]{Charlotte A. Mason}
\affiliation{Cosmic Dawn Center (DAWN)}
\affiliation{Niels Bohr Institute, University of Copenhagen, Jagtvej 128, 2200 Copenhagen N, Denmark}
\email{charlotte.mason@nbi.ku.dk}

\author[0000-0001-5940-338X]{Mengtao Tang}
\affiliation{Steward Observatory, University of Arizona, 933 N Cherry Ave, Tucson, AZ 85721, USA}
\email{}

\author[0000-0003-1432-7744]{Lily Whitler}
\affiliation{Steward Observatory, University of Arizona, 933 N Cherry Ave, Tucson, AZ 85721, USA}
\email{lwhitler@arizona.edu}

\author[0000-0002-4965-6524]{Ting-Yi Lu}
\affiliation{Cosmic Dawn Center (DAWN)}
\affiliation{Niels Bohr Institute, University of Copenhagen, Jagtvej 128, 2200 Copenhagen N, Denmark}
\email{tingyi-lu@nbi.ku.dk}

\author[0000-0001-8426-1141]{Michael W. Topping}
\affiliation{Steward Observatory, University of Arizona, 933 N Cherry Ave, Tucson, AZ 85721, USA}
\email{michaeltopping@arizona.edu}





\begin{abstract}

Ly$\alpha$ spectroscopy with {\it JWST} is opening a new window on the sizes of ionized bubbles through the reionization epoch. Theoretical expectations suggest typical bubble radii should be 0.6--1.5 pMpc at $z\simeq 7$, assuming neutral hydrogen fractions of the intergalactic medium in the range $\overline{x}_{\rm HI}$=0.5--0.7. Here we investigate this picture using {\it JWST} to characterize the environment and Ly$\alpha$ emission of 292 galaxies at $7.0<z<8.5$  across 5 fields spanning a comoving volume of $1.3\times10^6$~Mpc$^3$. If the reionization predictions are correct, we should see overdensities and strong Ly$\alpha$ emission clustered in redshift windows of d$z=0.04-0.08$ and angular scales of 5--11 arcmin.
We detect \lya{} emission in 36 out of 292 galaxies, including 
nine new \lya{} detections,  two of which  (in the UDS field)  show extremely large equivalent widths (EW = $200_{-78}^{+50}$~\AA{} and $284_{-75}^{+56}$~\AA{}).  We identify 13 significant (4--11$\times$) galaxy overdensities using redshifts from NIRCam grism and NIRSpec.  Strong Ly$\alpha$ emitters are almost uniformly found in the overdensities, with nearly all located between the center and back of the structures. The overdensities that host the strong Ly$\alpha$ emitters span typical line-of-sight distances (d$z\sim 0.14$) and angular scales ($\sim 8$ arcmin) that are comparable to the predicted bubble sizes at $z\simeq 7$. We discuss evidence that the EGS is mostly ionized along a 24 pMpc sightline at $z\simeq 7.0-7.6$, based on the presence of 3 overdense structures and 10 Ly$\alpha$ emitters in this volume, and find such a large ionized region would pose tension with standard reionization models.
\end{abstract}

\keywords{Early universe (435), High-redshift galaxies (734), Reionization (1383)}


\section{Introduction} \label{sec:intro}

The \lya{} emission of early star-forming galaxies provides an important probe of the reionization of intergalactic hydrogen (see \citealt{Dijkstra2014,Ouchi2020,Stark2025} for reviews).
Due to the large scattering cross section neutral hydrogen provides to Ly$\alpha$ photons, we expect  Ly$\alpha$ emission lines to be significantly 
attenuated at redshifts when the intergalactic medium (IGM) is significantly neutral.
For nearly two decades, concerted efforts have been made to statistically measure the \lya{} strength of continuum-selected galaxies at high redshifts \citep[e.g.,][]{Fontana2010,Stark2010,Ono2012,Pentericci2014,Hoag2019,Mason2019}.
Deep near infrared spectroscopy from the ground reveals that while strong \lya{} emitters are common at $z\sim5\mbox{--}6$, their fraction declines substantially at $z\gtrsm7$, suggesting increased attenuation to \lya{} photons at higher redshifts \citep[e.g.,][]{Ono2010,Stark2010,Treu2013,Schenker2014,Tilvi2014,Pentericci2018,Mason2019,
Bolan2022}.
If this attenuation is due to IGM, we may expect it to be significantly neutral at $z \gtrsim 7$ \citep[e.g.,][]{Caruana2014,Mason2018}, consistent with evidence from quasar absorption spectra \citep[e.g.,][]{Wang2020,Yang2020} and studies of the Cosmic Microwave Background \citep[e.g.][]{PlanckCollaboration2020}.

The launch of \jwst{} has rapidly advanced  \lya{} emission line investigations in the reionization era (see \citealt{Stark2025} for a review).
In addition to the absence of sky lines, \lya{} spectroscopy with \jwst{} has several advantages compared to observations from the ground.
With its improved sensitivity in the near infrared, \jwst{} enables meaningful \lya{} constraints at continuum magnitudes much fainter (by $\gtrsim3$ mag) than what was possible with ground-based $z>7$ spectroscopy \citep[e.g.,][]{Saxena2023,Chen2024}
The reliability of these measurements is further improved by confirmation of systemic redshifts through the detection of the continuum break or other emission lines.
These capabilities have been demonstrated by the \lya{} detection out to $z\sim13$ \citep[e.g.,][]{Witstok2024_z13} and growing number of galaxies with \lya{} measurements at $z\gtrsim7$ \citep[e.g.,][]{Tang2023,Nakane2023,Napolitano2024,Tang2024_nirspec,Jones2025,Kageura2025}.
Analyses with \jwst{} \lya{} measurements have led to new constraints on the reionization timeline, extending to very early epochs ($z\gtrsim10$; e.g., \citealt{Nakane2023,Napolitano2024,Tang2024_nirspec,Jones2025,Kageura2025}).

Attention has also been focused on interpreting the handful of detections of intense \lya{} emission lines at $z>7$ \citep[e.g.,][]{Tang2024_nirspec}.
The presence of strong \lya{} in a significantly neutral IGM may be explained if the host  galaxy resides in large ionized bubbles, allowing the \lya{} photons to  cosmologically redshift significantly before encountering the first patch of neutral hydrogen \citep[e.g.,][]{Wyithe2005,Furlanetto2005,Weinberger2018,Barkana2004,Iliev2006,Dayal2018,Weinberger2018}. If the bubbles are large, Ly$\alpha$ may redshift far enough into the damping wing where the opacity is greatly reduced.
These large ionized regions are predicted to first form around overdensities of galaxies  \citep[e.g.,][]{Furlanetto2004,Mesinger2007,Qin2021}.
Searches for early ionized structures have started shortly after the first detections of \lya{} at $z>7$ from the ground, but their characterization has been challenging due to difficulties in \lya{} observations from the ground and large uncertainties due to photometric uncertainties \citep[e.g.,][]{Castellano2016,Tilvi2020,Hu2021,Endsley2022_overdensity,Jung2022,Larson2022,Leonova2022}.

\jwst{} observations have significantly improved our ability to investigate the early ionized structures.
During the first Cycle of {\it JWST}, NIRSpec spectroscopy has confirmed several very strong LAEs at $z>7$ with rest frame \lya{} equivalent widths (EW) exceeding 100~\AA{} \citep[e.g.,][]{Saxena2023,Chen2024}.
The measured large \lya{} EWs suggest significant \lya{} transmission through the IGM, which may be expected if these galaxies reside in large ($\sim$pMpc) ionized bubbles.
Measurements of systemic redshifts for other galaxies in these fields have taken the first steps toward quantifying the environment around the Ly$\alpha$ emitters, identifying several potential large scale overdensities \citep[e.g.,][]{Tang2023,Chen2024,Napolitano2024,Tang2024_nirspec,Whitler2024,Witstok2024,Witstok2025_z8}.

After three years of \jwst{} operations, the database of NIRSpec spectra targeting $z>7$ galaxies has grown substantially across five deep extragalactic fields: UDS, EGS, GOODS-S, GOODS-N, and Abell 2744.
In this work, we conduct a systematic study of the impact of large scale environment on \lya{} emission at $z=7.0\mbox{--}8.5$, leveraging improved sample statistics and environmental characterization compared to earlier studies. 
We investigate the distribution of galaxy overdensities and the relative positions of \lya{} detections in each field.
We discuss the current constraints on the physical scales of the ionized sightlines and how we may fulfill the potential of \jwst{} in characterizing large ionized bubbles in the early universe.

\begin{table*}[ht!]
    \centering
    \caption{Summary of NIRSpec/MSA observations utilized in this work. We list the programs, PIs, the type of NIRSpec spectra where \lya{} is covered, and the corresponding references in each field.}
    \begin{tabular}{c|cccc}
    \hline
    \hline
    Field        & Program           & PI(s)                        & \lya{} Spectra      & References \\
    \hline
    UDS          & RUBIES (GO 4233)  & A. de Graaff                 & Prism               & [1]     \\
                 & CAPERS (GO 6368)  & M. Dickinson                 & Prism               & [2,3]   \\
    \hline
    EGS          & CEERS (ERS 1345)  & S. Finkelstein               & G140M/F100LP, Prism & [4]     \\
                 & DDT 2750          & P. Arrabal Haro              & Prism               & [4,5,6] \\
                 & RUBIES (GO 4233)  & A. de Graaff                 & Prism               & [1]     \\
                 & GO 4287           & C. Mason \& D. Stark         & G140H/F100LP        & [7]     \\
                 & CAPERS (GO 6368)  & M. Dickinson                 & Prism               & [2,3]   \\
    \hline
    GOODS-S      & JADES (GTO 1180)  & D. Eisenstein                & G140M/F070LP, Prism & [8,9]   \\
                 & JADES (GTO 1210)  & N. Lützgendorf               & G140M/F070LP, Prism & [8,10]  \\
                 & JADES (GTO 1286)  & N. Lützgendorf               & G140M/F070LP, Prism & [8,9]   \\
                 & JADES (GTO 1287)  & K. Isaak                     & G140M/F070LP, Prism & [8,9]   \\
                 & JADES (GTO 3215)  & D. Eisenstein \& R. Maiolino & G140M/F070LP, Prism & [11]    \\
    \hline
    GOODS-N      & JADES (GTO 1181)  & D. Eisenstein                & G140M/F070LP, Prism & [8,9]   \\
    \hline
    Abell 2744   & GLASS (ERS 1324)  & T. Treu                      & G140H/F100LP	      & [12,13] \\
                 & UNCOVER (GO 2561) & I. Labbé \& R. Bezanson      & Prism               & [14,15] \\
                 & DDT 2756          & W. Chen                      & Prism               & [16]    \\
                 & GO 3073           & M. Castellano                & Prism               & [17]    \\
    \hline
    \end{tabular}
    \tablerefs{
    [1] \cite{deGraaff2024},
    [2] Dickinson in prep.,
    [3] \cite{Kokorev2025},
    [4] \cite{Finkelstein2025},
    [5] \cite{ArrabalHaro2023_nat},
    [6] \cite{ArrabalHaro2023},
    [7] Whitler in prep.,
    [8] \cite{Eisenstein2023_JADES},
    [9] \cite{D'Eugenio2024},
    [10] \cite{Bunker2024},
    [11] \cite{Eisenstein2023_JOF},
    [12] \cite{Treu2022},
    [13] \cite{Mascia2024},
    [14] \cite{Bezanson2024},
    [15] \cite{Price2024},
    [16] \cite{Roberts-Borsani2023},
    [17] \cite{Castellano2024}.
}
    \label{tab:nirspec_data}
\end{table*}

The organization of this paper is as follows.
In \S~\ref{sec:data_sample}, we describe the sample of $z=7.0\mbox{--}8.5$ galaxies from the NIRSpec data.
We present new \lya{} detections in \S~\ref{sec:newLAEs}.
We then characterize the spatial distribution of galaxies and discuss where the \lya{} emitters are located in \S~\ref{sec:overden}.
In \S~\ref{sec:results_lya}, we statistically quantify the environmental effect on \lya{} transmission.
Throughout this paper, we adopt a flat $\Lambda$CDM cosmology with $H_0$ = 70 km~s$^{-1}$~Mpc$^{-1}$, $\Omega_{\rm m} = 0.3$, and $\Omega_\Lambda = 0.7$.
All magnitudes are measured in the AB system \citep{Oke1983}, and the emission line equivalent widths are calculated in the rest frame.

\section{Spectroscopic Data and Sample}\label{sec:data_sample}

\begin{figure*}[t!]
    \centering
    \includegraphics[width=1\textwidth]{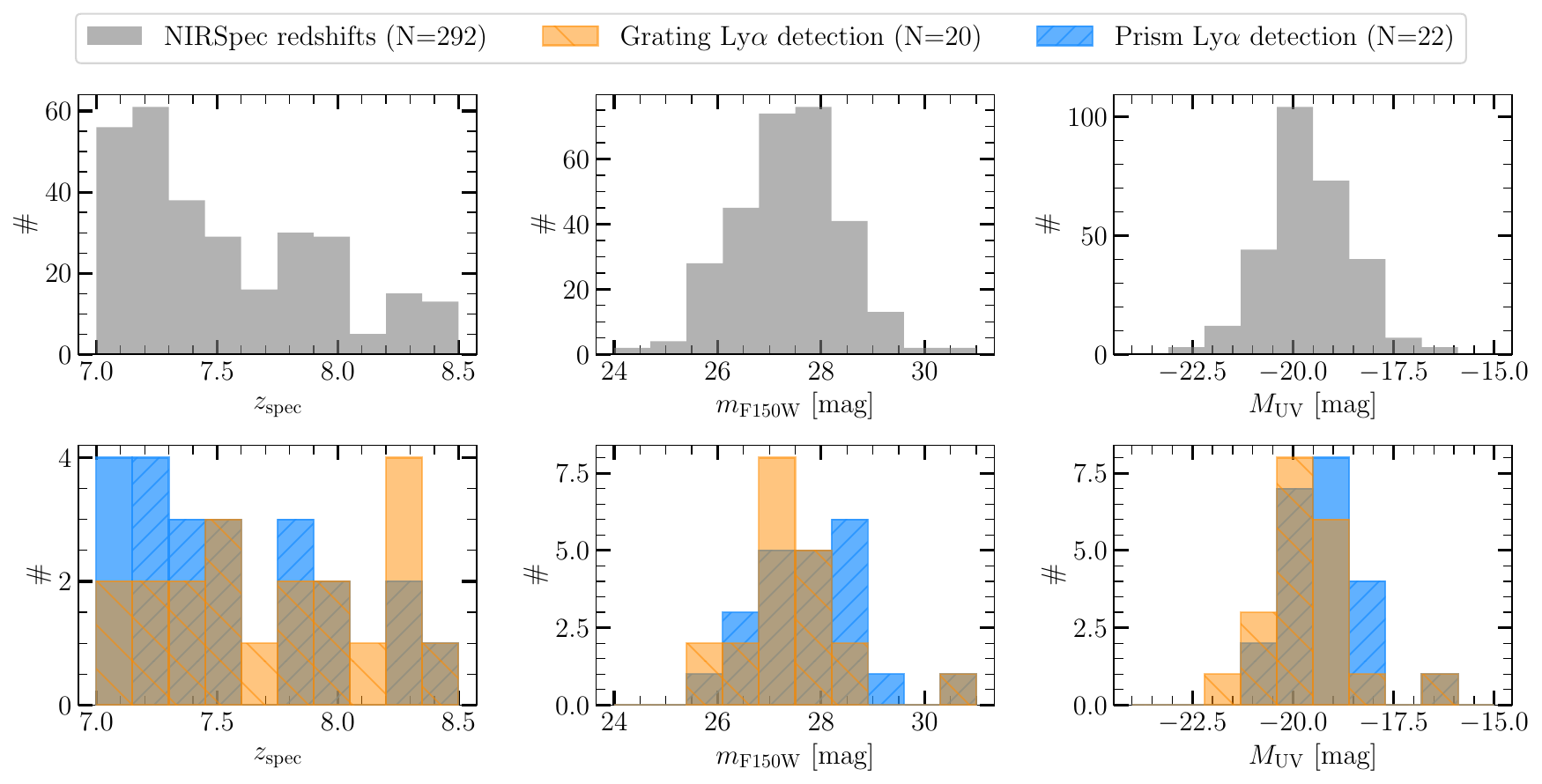}
    \caption{Histograms of spectroscopic redshifts, NIRCam F150W apparent magnitudes, and absolute UV magnitudes (\muv{}) for the galaxies analyzed in this work.
    Our sample includes in total 292 $z=7.0\mbox{--}8.5$ galaxies (top row), 20 (22) of which show \lya{} detections with grating (prism) spectra (bottom row).
    In general, we find galaxies with \lya{} detections show a similar redshift and magnitude distribution as the full sample. }
    \label{fig:sample_basic_hist}
\end{figure*}

\begin{figure*}
    \centering
    \includegraphics[width=1\textwidth]{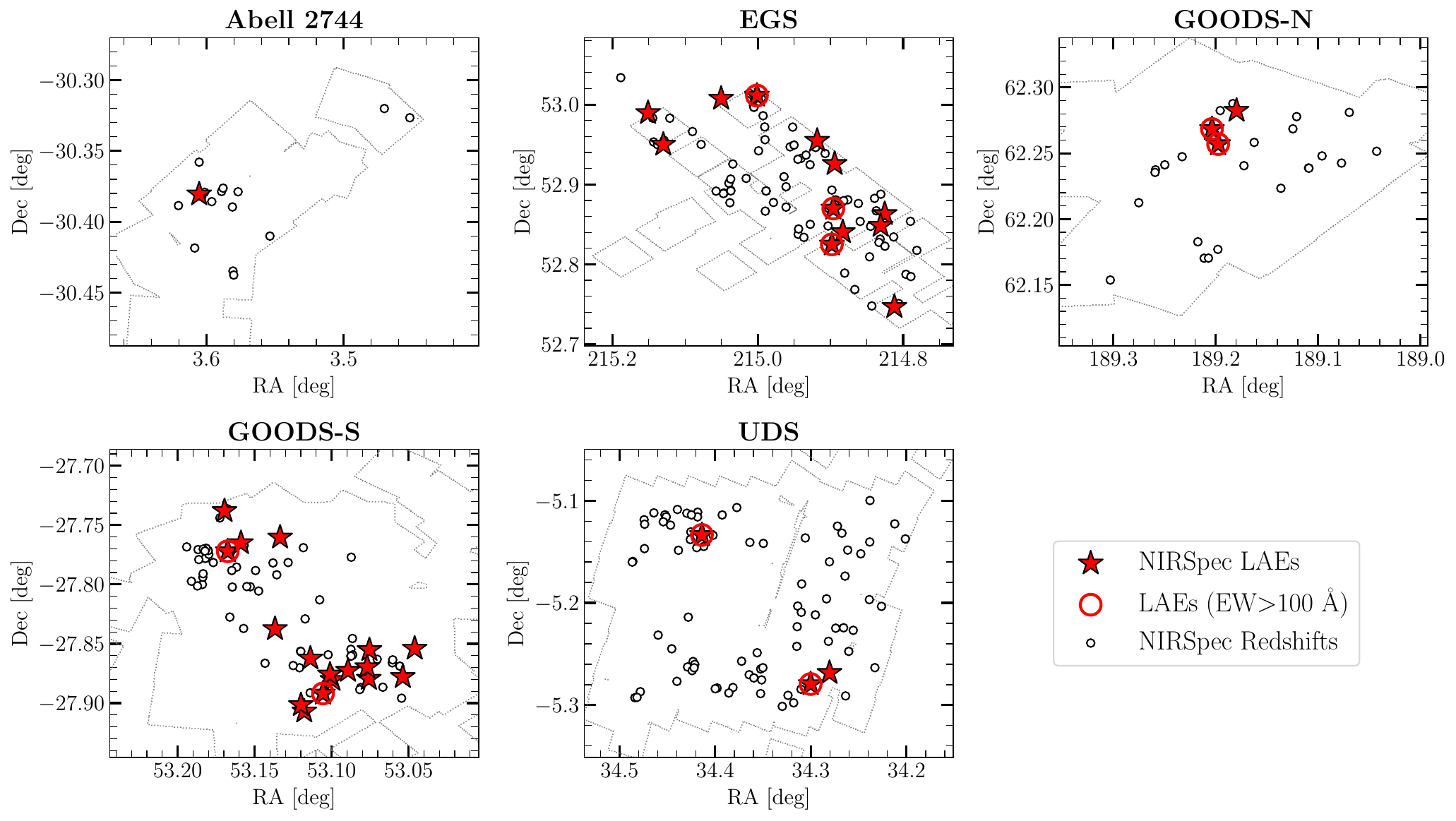}
    \caption{Spatial distribution of the \lya{} emitting galaxies (red stars) across our five survey fields at $7.0<z<8.5$.
    Those with extremely high equivalent widths (EW \lya{} $>$ 100 \AA{}) are highlighted with red circles.
    We additionally show the full sample of the NIRSpec targeted galaxies (black open circles) over the same redshift range for comparison.}
    \label{fig:2dmaps}
\end{figure*}

\subsection{\jwst{}/NIRSpec Spectroscopy}\label{subsec:data_nirspec}

Our work builds upon the \jwst{}/NIRSpec database compiled in Chen et al. in preparation and \cite{Tang2024_nirspec}, which includes public NIRSpec observations in five extragalactic fields: Abell 2744, Extended Groth Strip (EGS), Great Observatories Origins Deep Survey North and South (GOODS-N and GOODS-S), and the UltraDeep Survey (UDS).
This database incorporates NIRSpec spectra taken with multiobject spectroscopy (MOS) mode using the microshutter assembly (MSA;  \citealt{Ferruit2022}) through Cycles 1 to 3, including the most recent data in the EGS field taken in March 2025 as part of The CANDELS-Area Prism Epoch of Reionization Survey (CAPERS; GO 6368; PI: M. Dickinson) programs \citep{Dickinson2024}.
We summarize all the observations and the associated NIRSpec programs in Table~\ref{tab:nirspec_data}, and we refer the reader to the references listed there for more details.

Our goal is to investigate the \lya{} properties of $z\sim7.0\mbox{--}8.5$ galaxies, requiring observations with either the low-resolution ($R\sim100$) prism, median resolution ($R\sim1000$) grating G140M, and high resolution ($R\sim2700$) grating G140H.
Our current dataset includes 67 pointings taken with prism spread over all five fields, with a median exposure time of 1.68 hr.
An additional 25 pointings taken with F070LP/G140M are also available in the GOODS-N, GOODS-S fields (median exposure time 2.41 hr), and 6 pointings with G140M/F100LP in the EGS field (exposure time 0.86 hr).
Both grating/filter pairs provide the necessary coverage of \lya{} emission lines at our redshift of interest.
We further include the 4 pointings obtained using G140H/F100LP available in Abell 2744 and EGS fields, and the exposure times range from 3.9--4.9 hr.
For the grating pointings, we will also consider available observations taken at longer wavelengths (i.e., G235M/F170LP, G395M/F290LP, G235H/F170LP, G395H/F290LP) for identifying galaxy systematic redshifts.

All NIRSpec MOS spectra were reduced uniformly following the procedures detailed in \cite{Topping2024}, which we briefly summarize below.
We utilized the standard JWST data reduction pipeline\footnote{\url{https://github.com/spacetelescope/jwst}} \citep{Bushouse2024} in addition to custom routines.
We started with the raw, uncalibrated images, where we flagged cosmic rays, subtracted ‘snowballs’ and ‘showers’ artifacts, performed ramp fitting, and corrected for $1/f$ noise
From each of the resulting rate images, we created cutouts of 2D spectrum traced out by individual targeted objects, then applied flat-field correction, wavelength solution, and absolute photometric calibration.
In this step, we assumed a point source for wavelength-dependent slit loss correction, as the majority of the microshutters sample compact galaxies (or sub-regions within them). 
We will come back to comment on this in more detail in \S~3.
Finally, for each target, we background subtracted all 2D cutouts, which are then combined and interpolated onto a common wavelength grid to obtain the final 2D spectra.
Following \cite{Tang2024_nirspec}, we extracted the final 1D spectrum for each source from the reduced 2D spectrum with a boxcar window, where the width was set to match the continuum or the emission line profile along the spatial direction (median $\sim$5 pixels).

\subsection{Sample Selection and Galaxy Catalogs}\label{subsec:sample}

We visually inspect each 2D and 1D spectrum in search of objects at $7.0<z<8.5$.
To determine the redshift of each object, we utilize modules in {\sc msaexp} \citep{Brammer_msaexp_NIRSpec_analyis_2022} that fit the full spectra with {\sc Eazy-py} \citep{Brammer2008,Brammer_eazy-py_2021}.
To ensure robust redshift identification, we only consider objects with at least two emission lines (often \ha{} and [\oiii{}]5008) detected at $\rm S/N > 3$ or the presence of \lya{} break.
To focus on the \lya{} properties in star-forming galaxies, we do not select sources with significantly broader Balmer emission lines (\ha{} or \hb{}, FWHM $>$ 1500~\kms{}) than [\oiii{}] indicative of the presence of active galactic nuclei (see e.g., \citealt{Harikane2023_AGN,Kocevski2023,Greene2024,Maiolino2024}).
From the full NIRSpec dataset, we end up with a final sample of 292 galaxies at $7.0<z<8.5$ (median redshift 7.43).
The sample size is significantly (more than $\times$3) larger compared to the previous ones at $7.0<z<8.5$ \citep[e.g.,][]{Tang2024_nirspec,Kageura2025}, providing us the statistics required to characterize the \lya{} emission strength at these redshifts. 
In the Abell 2744 field, we also consider the spectroscopically confirmed galaxies over this redshift range from the NIRCam F356W grism observations taken by the All the Little Things (ALT; GO 3516; PI J. Matthee \& R. Naidu; \citealt{Naidu2024}) program.
Across a comparable footprint (30~arcmin$^{2}$) as the NIRSpec observations, ALT confirms four galaxies at $z=$7.0--8.5 via detection of the \hg{} emission line, one of which is new and has not been targeted by NIRSpec.
We will include this galaxy to increase the statistics for characterizing the spatial distribution of galaxies in the Abell 2744 field in \S~\ref{sec:overden}, but the rest of our analyses will be primarily focused on the galaxies from the NIRSpec sample.

We utilize available \hst{}/ACS+\jwst{}/NIRCam observations in the five fields to derive the photometric properties necessary for interpreting the \lya{} emission in our spectroscopic sample.
We will also use these imaging data to identify photometric candidates at similar redshifts to the spectroscopic sample in \S~\ref{subsec:photoz}.
All NIRCam data are collected and reduced homogeneously following the description in \cite{Endsley2024_a2744}.
We also included \hst{}/ACS imaging assembled and reduced with {\textsc{Grizli}} \citep{Brammer2022} as part of the Complete Hubble Archive for Galaxy Evolution project (\citealt{Kokorev2022}; Kokorev et al. in preparation) and the ACS mosaics available from the Dawn \jwst{} Archive. 
We measure Kron photometry in each of the available ACS and NIRCam filters for sources in our NIRSpec sample.
We show the distribution of their NIRCam F150W magnitudes in the bottom middle panel of Figure~\ref{fig:sample_basic_hist}, which ranges from m$_{\rm{AB}}=26.4$ to 28.4 (inner 68\% range, same below) with a median of 27.4 mag.
We also derive the absolute UV magnitudes at rest-frame wavelength of 1500~\AA{} through a power law fitting ($f_\nu\propto \lambda^{-\alpha}$) to the continuum flux densities (in $f_\nu$) in filters covering from rest-frame wavelengths of 1250 \AA{} to 2600 \AA{}.
We correct for lensing magnification with the \cite{Furtak2023_lesning_model} lensing model for sources in the Abell 2744 field (median magnification $\mu=1.62$).
The resulting absolute UV magnitudes for our sample span from \muv{}=$-20.6$ to $-18.9$ with a median of $-19.7$ mag.

\subsection{\lya{} Emission Measurements}\label{subsec:lya_measurement}

\begin{figure*}
    \centering
    \includegraphics[width=\textwidth]{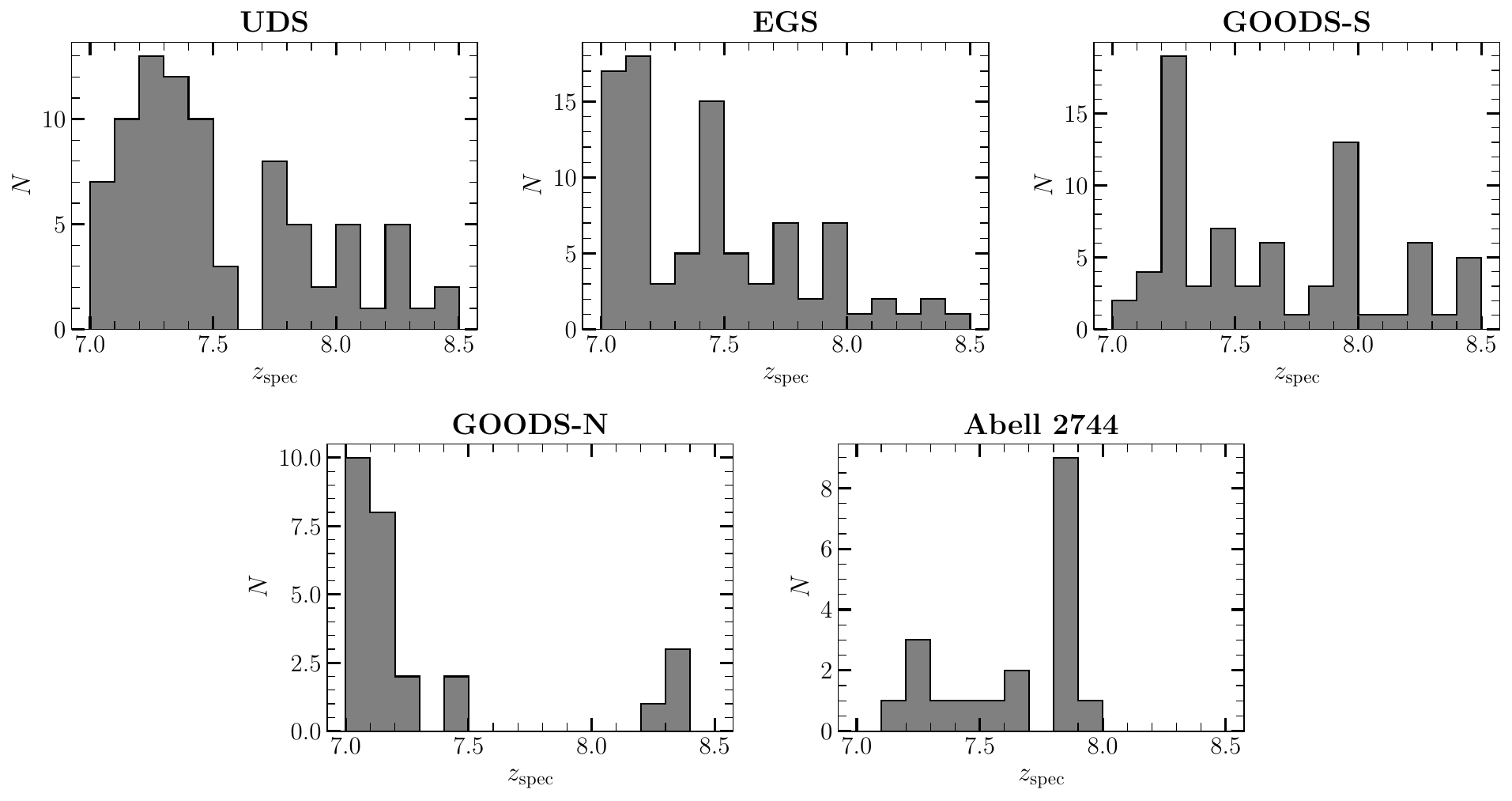}
    \caption{Histogram of NIRSpec redshifts in the UDS, EGS, GOODS-S, GOODS-N, and Abell 2744 fields. Several of the peaks correspond to galaxy overdensities. Additional overdensities become clear when the redshift distribution is investigated in sub-regions of the fields.}
    
    \label{fig:zhist_nirspec}
\end{figure*}

\begin{figure*}[t!]
    \centering
    \includegraphics[width=0.49\textwidth]{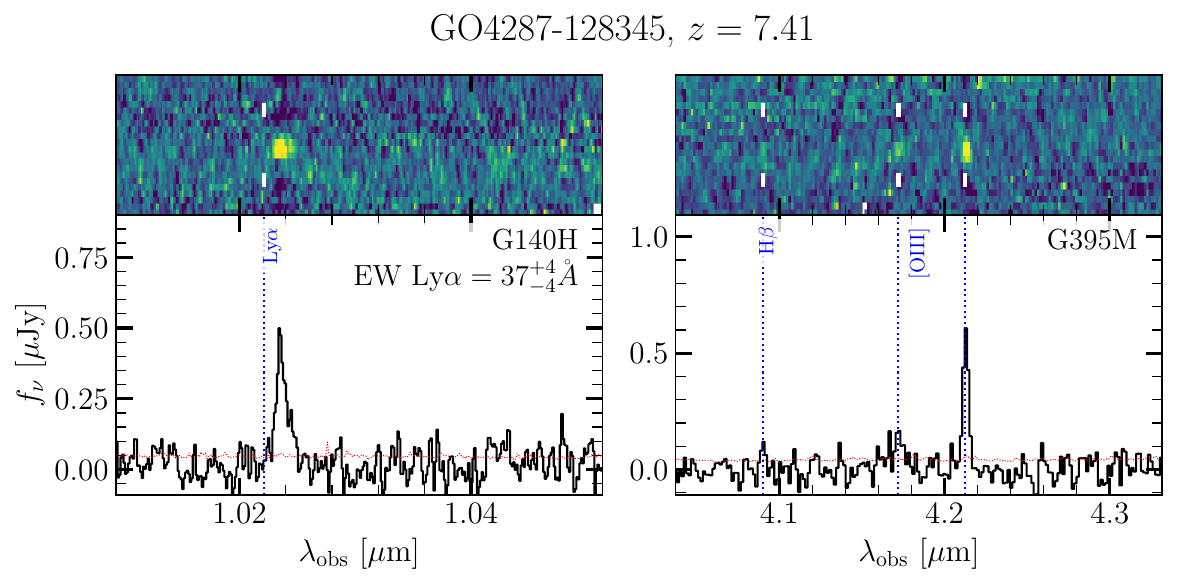}
    \includegraphics[width=0.49\textwidth]{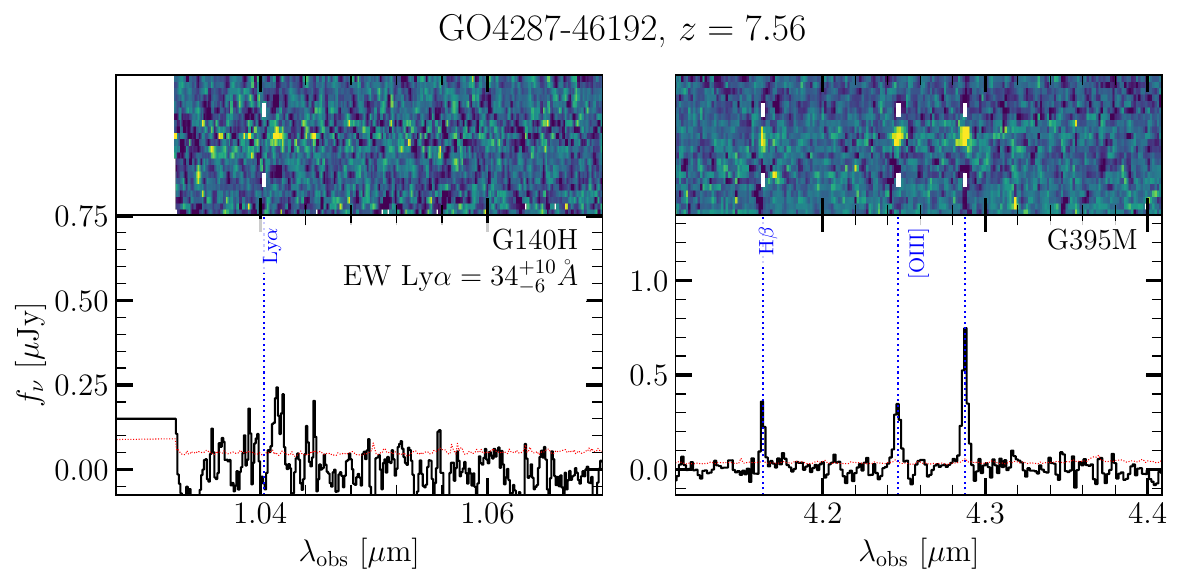}
    \includegraphics[width=0.49\textwidth]{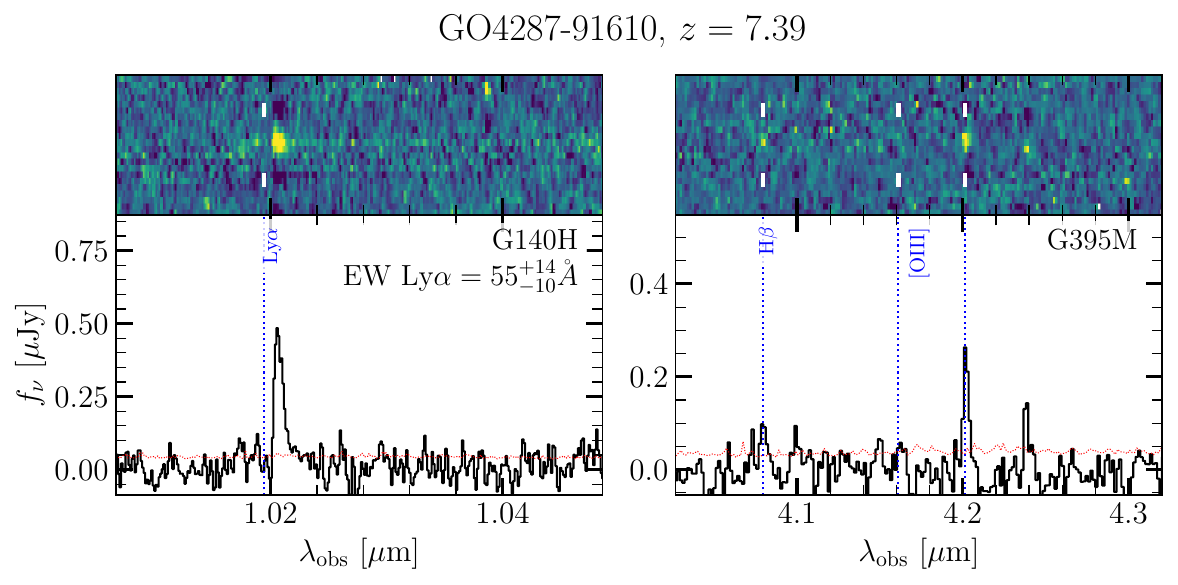}
    \includegraphics[width=0.49\textwidth]{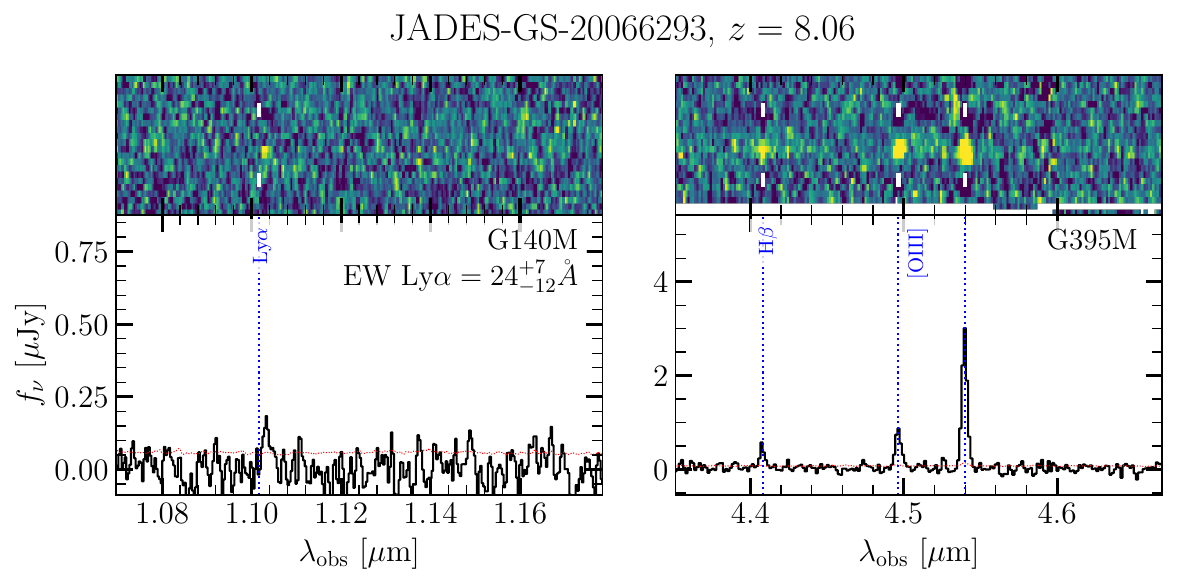}
    \includegraphics[width=0.49\textwidth]{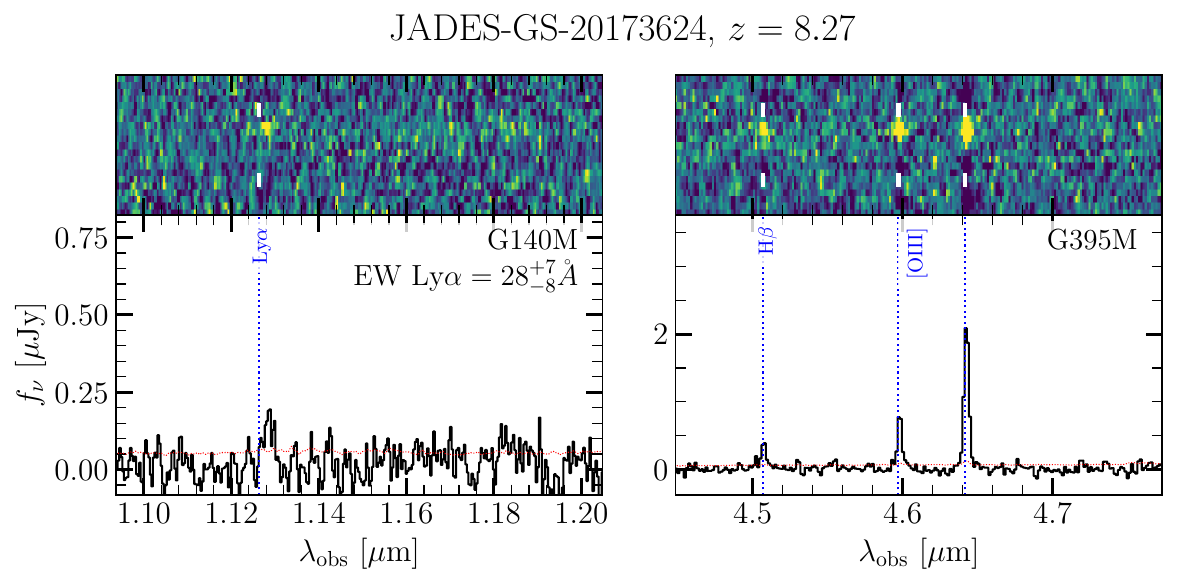}
    \caption{New \lya{} emitting galaxies at $z\sim7.0\mbox{--}8.5$ detected with NIRSpec medium to high resolution grating spectra.
    Both 1D and 2D spectra are presented.
    For each galaxy, we show both the \lya{} detection in the left panel and the optical emission lines ([\oiii{}] and \hb{}) in the right panel.
    Blue vertical dotted lines indicate the expected position of each line given the systemic redshift.
    }
    \label{fig:newLAE_grating}
\end{figure*}

We visually identify \lya{} emission at $z=7.0\mbox{--}8.5$ from the 2D and 1D spectra of our sample by searching for line features located at the expected wavelength given their systemic redshifts.
We require a minimum S/N ratio of 3 for detection, and we also check individual exposures of each source with a detection to avoid the inclusion of artifacts (i.e., hot pixels, cosmic rays). 
This yields a total number of 36 unique \lya{} detections, with 20 \lya{} detections from grating spectra and 22 from prism spectra (6 detected in both grating and prism spectra).
We estimate the S/N of the prism detections to range from 3.0--44.3 (median 5.4), and those of the grating detection to range from 3.0--30.5 (median 5.9).
We identify 9 new \lya{} detections over this redshift range, which we will discuss individually in more detail in \S~\ref{sec:newLAEs}.

We adopt slightly different methods for Ly$\alpha$ flux and EW measurements in grating and prism spectra. 
For each grating spectrum, we follow \cite{Tang2024_nirspec} to estimate and subtract the continuum underlying the \lya{} emission before deriving the line flux.  
We calculate the continuum as the average flux density (in $f_\nu$) in the spectra at rest-frame wavelengths of $\lambda=1300$--1400 \AA{}, which is chosen to minimize the impact from damped Ly$\alpha$ absorption on the continuum estimation \citep[e.g.,][]{Heintz2023}. We compute the continuum in $f_\lambda$ at the rest-wavelength of Ly$\alpha$ assuming a flat spectrum in $f_\nu$, which is typical for the galaxies in our sample. 
For each \lya{} detection, we then compute the  flux by integrating the continuum-subtracted spectrum over a wavelength window of 10~\AA{} ($2500$~\kms{} in velocity space) in the rest frame centered at the \lya{} wavelength.
We perturb the observed spectrum according to its errors and repeat the measurements 1000 times to derive the median and uncertainty of the \lya{} flux.
For \lya{} non-detections, we derive 3$\sigma$ upper limits using the flux uncertainties computed via the same method. We compute EWs and EW limits using the derived continuum flux density at the Ly$\alpha$ wavelength.

The \lya{} fluxes and EWs from the prism spectra are derived using a similar method. To compute the continuum level, we fit a power-law function ($f_\lambda\propto\lambda^{\beta}$) to the spectra over rest-frame wavelengths of 1300--1700 \AA{} with fixed slope $\beta=-2$ (effectively assuming a flat continuum in $f_\nu$) and extrapolate it to the wavelength of \lya{}.
To derive the prism \lya{} fluxes, we integrate the continuum-subtracted line profile over rest-frame wavelength 1170--1270 \AA{} around the line center.
This yields the continuum observed in a consistent aperture as the \lya{} fluxes, with which we compute the line EWs.
In 5 (of  22) \lya{} detections in the prism spectra, the observed continuum at wavelengths longer than the \lya{} shows evidence of damped \lya{} absorption.
For these 5 sources, we use a refined version of the continuum when computing the line flux. In particular, we derive the local continuum level near \lya{} by fitting the spectrum with a linear function (in $f_\lambda$) over rest-frame wavelengths of 1280--1500 \AA{}. We then subtract this function from the spectrum to compute the line flux.

Due to the low spectral resolution of the prism, the \lya{} emission line will be blended with the \lya{} break, causing the measured line fluxes to be lower than their true values.
We correct for this flux loss using the approach presented in \cite{Chen2024} (also see \citealt{Jones2024}),  simulating  prism spectra to quantify the impact of spectral blending on the recovered Ly$\alpha$ fluxes. For each source in our sample, we generate mock  spectra with a range of intrinsic EWs from 1 to 1000 \AA{}. We adopt the average Ly$\alpha$ line profile measured at $z\simeq 5-6$ from \cite{Tang2024_z56}, although our results are not very sensitive to this choice given the coarse resolution of the prism. We then convert the spectrum to the resolution of the prism and compute the line fluxes and EWs as we describe above. This allows us to compute the mapping between the intrinsic EW and that which is observed. In general, we find corrections tend to be 3.3$\times$ at moderate intrinsic EW (30~\AA). For the sources with the highest intrinsic EW  (100~\AA), the corrections are smaller (1.35$\times$). For each source in our sample, we compute the intrinsic Ly$\alpha$ EW that maps to our observed value. Using the small sample of 5 sources with \lya{} confidently detected in both prism and grating spectra and not impacted by damped \lya{} absorption, we find that the IGM-corrected prism EWs agree with those measured from grating within 1$\sigma$ (median difference 5\%). 
In contrast, if we did not apply these corrections, the prism and grating spectra would have a median offset of 35\%, with prism systematically lower in EW as expected.

We will also use the \lya{} escape fraction to interpret the \lya{} transmission through IGM for the \lya{} emitters.
We derive the \lya{} escape fractions as the ratio between the observed and the intrinsic \lya{} fluxes, following the previous works \citep[e.g.,][]{Chen2024,Tang2024_nirspec}. 
We calculate the intrinsic \lya{} flux using the \hb{} emission line, as \ha{} has shifted out of NIRSpec wavelength coverage at $z\geq7$.
We detect \hb{} with S/N$>3$ for 31 of 35 \lya{} emitters, for which we compute the fluxes through Gaussian profile fitting.
We do not attempt to correct the \hb{} flux for dust, as the \hg{} emission line in the individual spectra is often not detected to allow for measurements of the Balmer decrement, and galaxies at these redshifts are expected to be relatively dust free (estimated from their composite spectra; e.g., \citealt{Tang2023}).
Assuming case B recombination ($T=10^4$~K), we expect an intrinsic ratio between \lya{}/\hb{} ratio of 25.0 \citep{Osterbrock2006}.
For the 36 galaxies with both \lya{} and \hb{} detected, we measure a median \lya{} escape fraction of 0.22 (inner 68\% range 0.13--0.52).
We also place 3$\sigma$ upper limits for escape fractions when \lya{} is not detected, with the median upper limit of 0.34.

\subsection{Photometric selection}\label{subsec:photoz}

In addition to investigating the distribution of spectroscopically confirmed galaxies, we also consider the spatial distribution of photometric candidates. Here we focus on fields that do not have published NIRCam grism coverage. In these cases, the measurement of spectroscopic overdensities relies on NIRSpec follow-up, which depends strongly on survey selection functions and tends to be significantly incomplete at a fixed magnitude or emission line flux. By mapping the distribution of photometric candidates across individual fields, we can identify sightlines that appear strongly overdense, which we can compare against possible overdense regions identified from the distribution of NIRSpec redshifts. 

We will focus our photometric search on the UDS and EGS fields, as these are the two widest-area imaging datasets lacking published grism coverage.
We select galaxies over 300~arcmin$^2$ in the UDS and 116~arcmin$^2$ in the EGS.
The \hst{}/ACS+\jwst{}/NIRCam imaging and photometric catalogs are described in \S~\ref{subsec:sample}.
The catalog includes NIRCam photometry in seven broad band filters (F090W, F115W, F150W, F200W, F277W, F356W, and F444W) and at least 1 medium-band filter (F410M in the UDS and F410M+F480M in the EGS, with both filters not uniformly available across the fields), as well as ACS photometry in at least F606W and F814W.
For our targets, the photometric redshifts are constrained by the presence of the Ly$\alpha$ break in the F090W filter. Additional constraining power comes from the impact of strong rest-frame optical emission lines on medium and broad-band filters at 3-5$\mu$m. In particular, the [\oiii{}] and H$\beta$ emission lines are situated in the F410M filter at $z\simeq 7.0-7.6$, creating a strong flux excess relative to adjacent filters. At $z\simeq 7.6-8.5$, the [\oiii{}] and H$\beta$ emission lines are shifted out of F410M (but remain in the F444W filter), producing an excess in F444W relative to bluer NIRCam filters. In both cases, the emission line excesses help narrow the photometric redshift confidence intervals relative to what was possible with just the imprint of the Ly$\alpha$ break.

We will determine photometric redshift constraints using the \textsc{Eazy-py} package, the \textsc{Python} version of \eazy{} \citep{Brammer2008,Brammer_eazy-py_2021} and adopting the 
 \cite{hainline_2023_7996500}  templates that are designed to identify high-redshift galaxies. We allow the code to explore the redshift range $z=0.01-20$ in steps of $\Delta z=0.01$ to output both the best-fit redshift as well as the redshift probability $P(z)$.  We employ the \eazy{} redshift probability $P(z)$ for each source to identify galaxies likely at $z=7.0\mbox{--}8.5$. We conservatively only consider sources with  $P(7.0<z<8.5) > 50\%$.
We further require a S/N$>3$ in the F150W band (ensuring detection in the rest-UV continuum).
Following this selection, we visually inspect the images and the ACS+NIRCam SED of every object to remove those likely to be stars or artifacts, the latter including diffraction spikes, objects coincident with the detector edge, and residuals left over from the cosmic ray removal.
We are left with 262 galaxies in the EGS and 393 galaxies in the UDS. We cross-match this catalog with our spectroscopic database. We verify that there are no catastrophic outliers, with redshifts well outside of our adopted range ($z<6$). The majority of the 118 confirmed sources lie in the desired redshift range, with a small subset found just outside ($\delta z=0.2$) our redshift window. We will explore the spatial distribution of these galaxies in \S~\ref{sec:overden}, identifying photometric overdensities and comparing them to the distribution of spectroscopically-confirmed galaxies.

\subsection{SED Modeling}\label{subsec:sed}

We will consider the dependence of \lya{} emission on the physical properties of galaxies in our analysis below.
Therefore, we characterize and fit the spectral energy distributions of galaxies in our sample to derive constraints on their physical properties.
Following our previous works, we measure the available \hst{}/ACS and \jwst{}/NIRCam photometry for each source in the sample and fit the SED with the photoionization modeling code \beagle{} \citep{Chevallard2016}.
We adopt the \cite{Gutkin2016} models that self-consistently combine the most recent version of \cite{Bruzual2003} stellar population models with nebular emission computed from {\sc Cloudy} \citep{Ferland2013}.

Our fitting follows the setup described in our previous works in modeling the high-redshift NIRSpec confirmed galaxies \citep[e.g.,][]{Chen2024,Tang2024_nirspec}. 
We fix each galaxy to its spectroscopic redshift.
We fit only in filters redward of the observed \lya{} wavelength and adopt a 5\% uncertainty floor for the photometry.
For simplicity, we assume a constant star formation history (CSFH), allowing the galaxy age to vary between 1 Myr to the age of the Universe at the given redshift with a log-uniform prior. 
We note that this approach will likely underestimate the total stellar mass owing to outshining in the case of galaxies dominated by young stellar populations \cite[e.g.,][]{Tang2022,Tacchella2023,Whitler2023_cosmos}. 
As we are not focused on trends with the true stellar population age or stellar mass in this study, this will not significantly impact our results. 
We adopt the \cite{Chabrier2003} initial mass function with the upper-mass cutoff of 300 $M_\odot$, with log-uniform priors on the total stellar mass ($5\leq {\rm log}(M_*/M_\odot) \leq 12$).
We place log-uniform priors on stellar metallicity ($-2 \leq {\rm log} (Z/Z_\odot) \leq -0.24$), assuming the total interstellar (gas phase + dust) metallicity is kept the same as the stellar metallicity through a fixed dust-to-metal mass ratio of $\xi_{\rm d} = 0.3$.
The gas ionization parameter is assumed to vary from $-4 \leq {\rm log} U \leq -1$ with a uniform prior in logarithmic space.
The resulting models are then attenuated by the interstellar medium assuming the \citealt{Pei1992} SMC dust attenuation curve, with V-band optical depth varying$-3.0\leq {\rm log}(\tau_{\rm \scriptscriptstyle{V}})\leq0.7$, and the intergalactic medium using the model of \cite{Inoue2014}.

\begin{figure*}[ht!]
    \centering
    \includegraphics[width=0.49\textwidth]{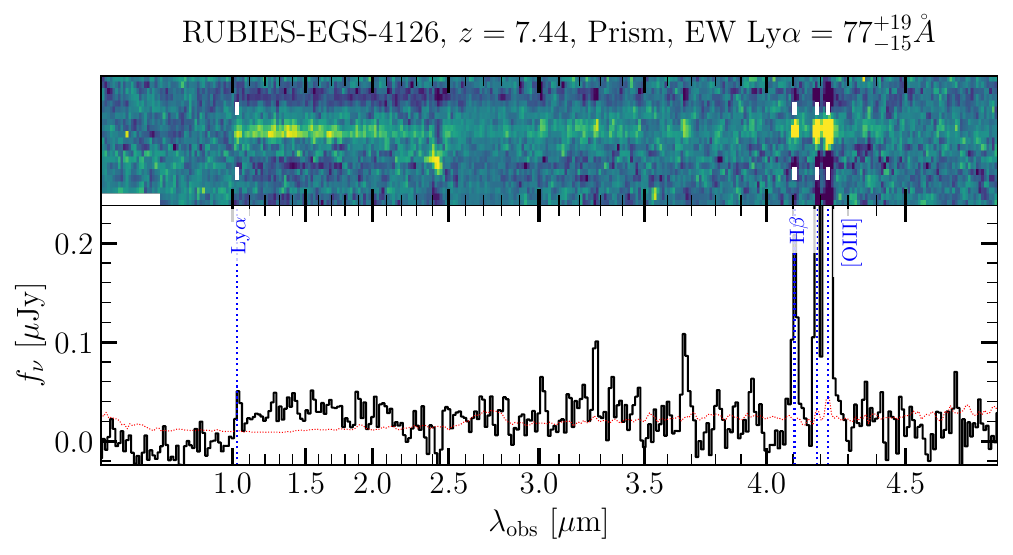}
    \includegraphics[width=0.49\textwidth]{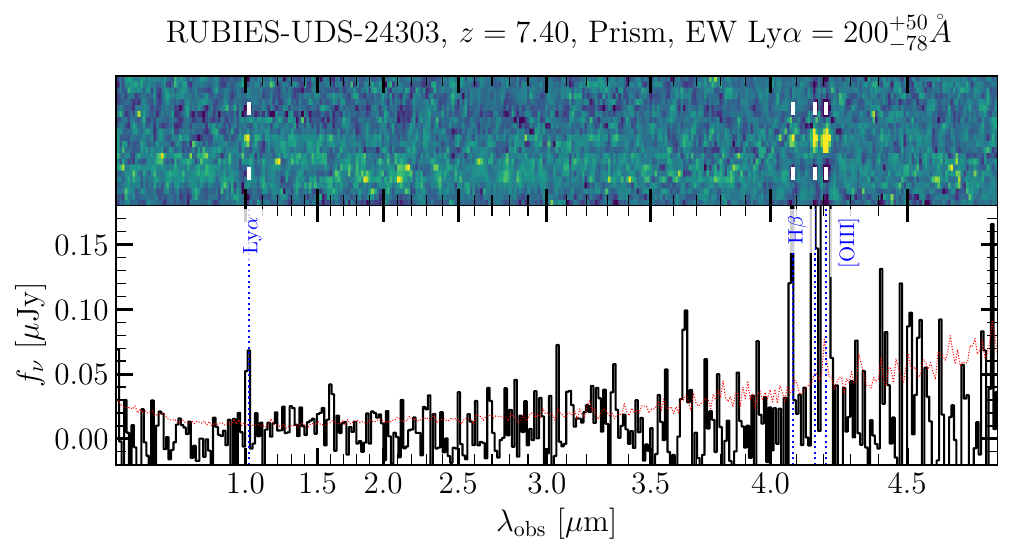}
    \includegraphics[width=0.49\textwidth]{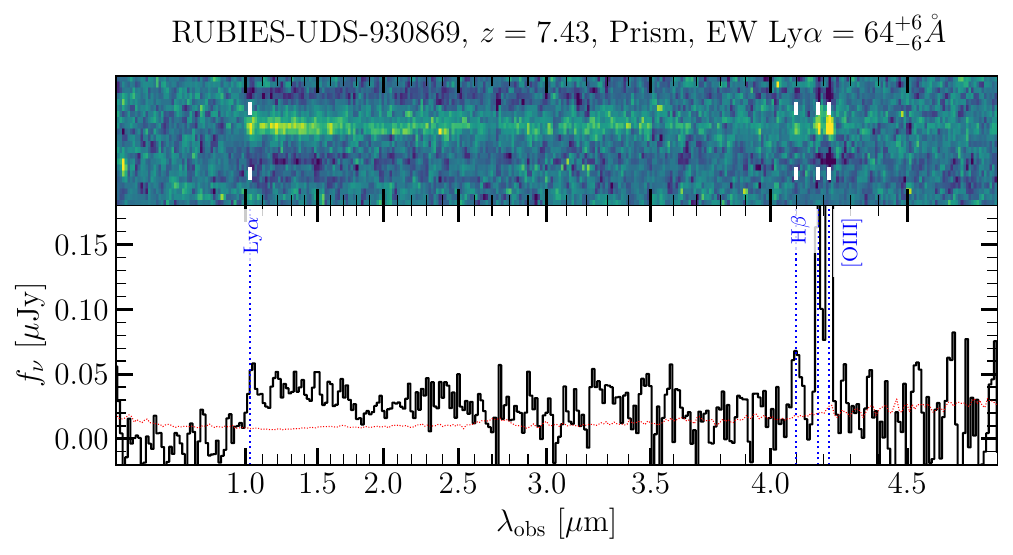}
    \includegraphics[width=0.49\textwidth]{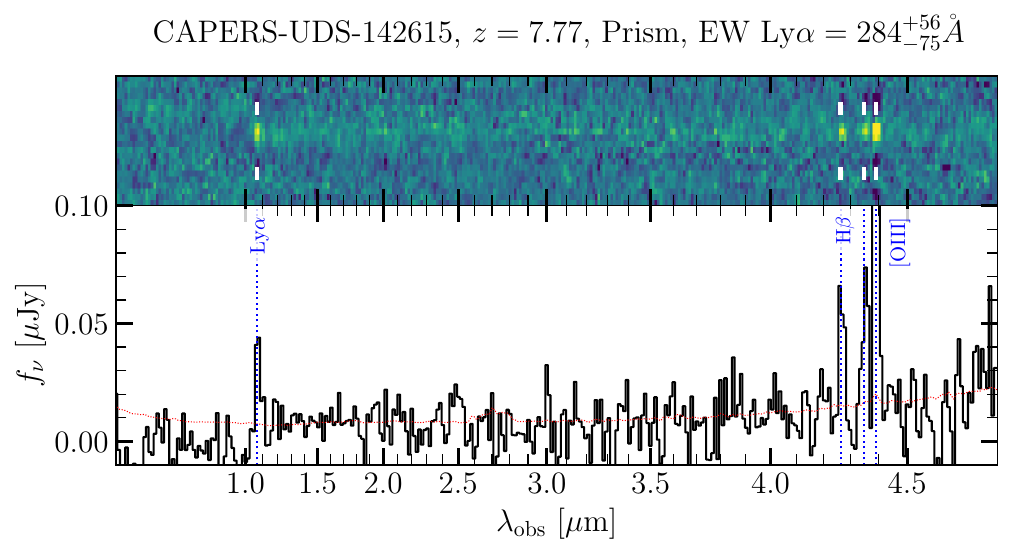}
    \caption{New \lya{} emitting galaxies at $z\sim7.0\mbox{--}8.5$ detected with NIRSpec prism spectra.
    We show both the 2D (top) and 1D spectrum (bottom) for each source, with detections of \lya{} and strong optical emission lines labeled in blue text.}
    \label{fig:newLAE_prism}
\end{figure*}

\section{New $z>7$ \lya{} Emission Line Detections}\label{sec:newLAEs}

Our sample includes new Ly$\alpha$ spectroscopy taken as part of several recent programs (CAPERS, RUBIES, GO 4287). In this Section we briefly present new detections of Ly$\alpha$ at $7.0<z<8.5$ from these programs. We also comment on the improved spectroscopic statistics that are now possible in each of our survey fields.
More details on the specific programs are in \S~\ref{subsec:data_nirspec}.

In the EGS, our current sample includes a total of 12 \lya{} detections, of which four are reported for the first time.
Three of the four are detected in F100LP/G140H grating spectra from the GO 4287 program  (GO4287-91610, GO4287-128345, and GO4287-46192; see Figure~\ref{fig:newLAE_grating}), and the fourth is discovered in a prism spectrum taken in the RUBIES program (RUBIES-EGS-4126; see Figure~\ref{fig:newLAE_prism}). The galaxies are of moderate luminosity (M$_{\rm{UV}}=-19.3$ to $-20.4$) with relatively strong \lya{} emission (EW ranging from 34 to 77~\AA{}).
In addition to the new \lya{} detections, the CAPERS, RUBIES, and GO 4287 observations have additionally confirmed the redshifts for 55 galaxies, more than doubling the number of confirmed sources over $z=7.0\mbox{--}8.5$.

Much of the spectroscopy considered in the UDS is relatively new, with 84 $7.0<z<8.5$ galaxies from  CAPERS and RUBIES. 
We present the 2D and 1D prism spectra of three new Ly$\alpha$ detections in Figure~\ref{fig:newLAE_prism}. The galaxies have absolute magnitudes ranging between M$_{\rm{UV}}$= $-18.9$ and $-19.3$. 
Two of these systems are found at similar redshifts ($z=7.40$ and 7.43), with one showing very large \lya{} EW (RUBIES-UDS-24303, EW = $200_{-78}^{+50}$~\AA{}) and the other also emitting moderately strong \lya{} emission (RUBIES-UDS-930869, EW = $64_{-6}^{+6}$~\AA{}).  We also measure very strong \lya{} for the third source: RUBIES-UDS-142615 (EW = $284_{-75}^{+56}$~\AA{}) at $z=7.77$.  

Finally, we also present two new \lya{} detections in the GOODS-S field observed by the GTO 1286 program: JADES-GS-20066292 at $z=8.06$, and JADES-GS-29173624 at $z=8.27$.
These two detections are not included in the previous compilation of  \lya{} emitters in the GTO 1286 dataset \citep[]{Jones2025}.
Using the F070LP/G140M grating spectra shown in Figure~\ref{fig:newLAE_grating}, we measure relatively small \lya{} EWs: $24_{-12}^{+7}$~\AA{} for JADES-GS-20066292 and $28_{-8}^{+7}$~\AA{} JADES-GS-29173624, which are the weakest among the new Ly$\alpha$ detections.

\section{Characterization of Overdensities }\label{sec:overden}

\begin{figure*}
    \centering
    \includegraphics[width=\textwidth]{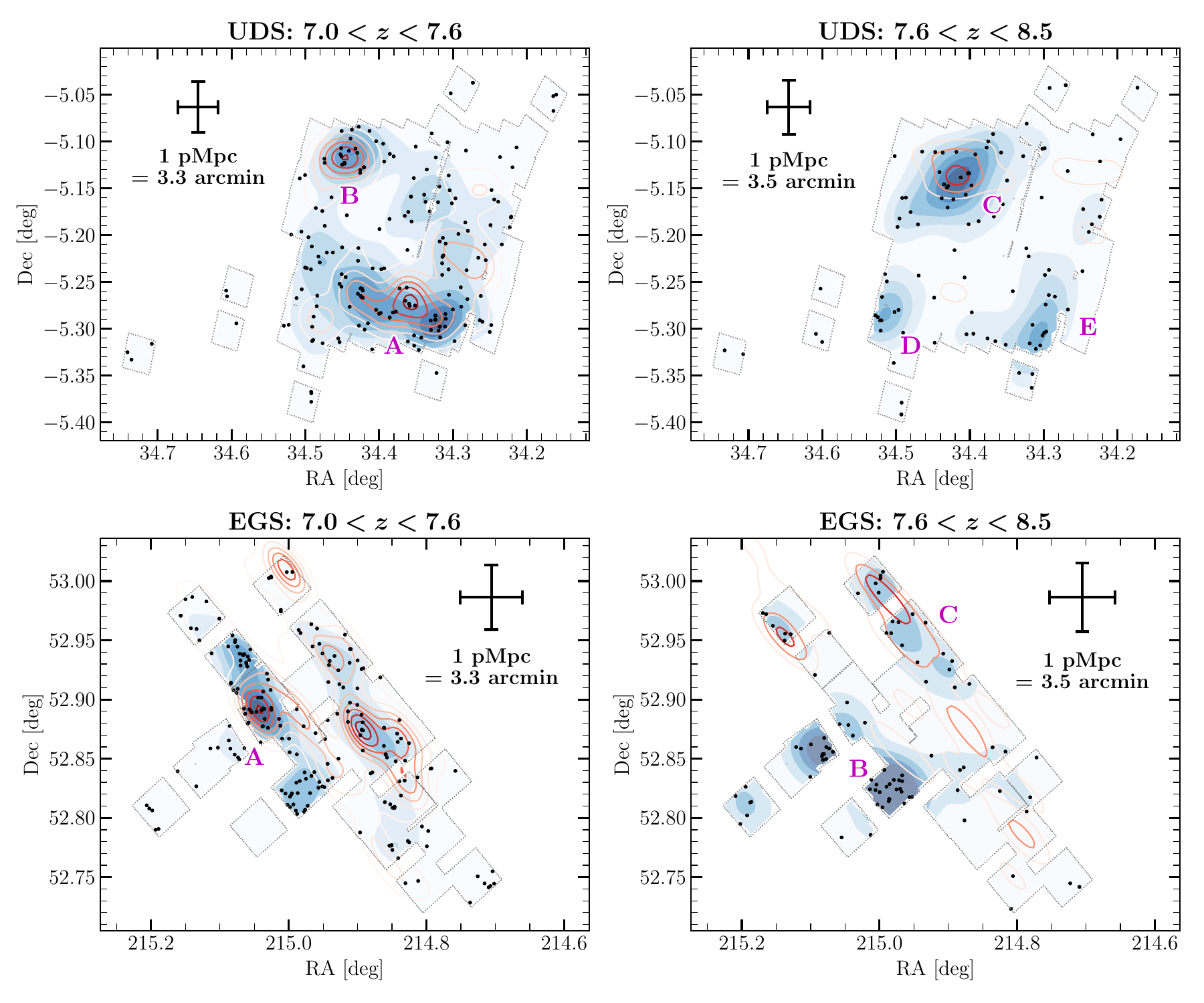}
    \caption{Spatial distribution of the photometrically-selected galaxies at $z=7.0\mbox{--}8.5$ in the UDS (top) and EGS (bottom) fields.
    We split the sample into two redshift bins: $7.0<z<7.6$ (left) and $7.6<z<8.5$ (right). The photometric targets are shown as black dots, while the blue shaded colors indicate the implied surface  densities (bluer colors correspond to higher surface  densities).
    We overplot the distribution of  NIRSpec-confirmed galaxies as red contours.
    }
    \label{fig:photoz}
\end{figure*}

In this section, we investigate the distribution of galaxies  across our five fields and describe the location of Ly$\alpha$ detections with respect to candidate overdense structures. We first describe techniques for identifying overdensities, highlighting different methods for fields with grism and those with NIRSpec. Then we provide an overview of the distribution of galaxies over $7.0<z<8.5$ in each field.

\begin{figure*}[t!]
    \centering
    \begin{minipage}[c]{0.52\textwidth}
        \includegraphics[width=\linewidth]{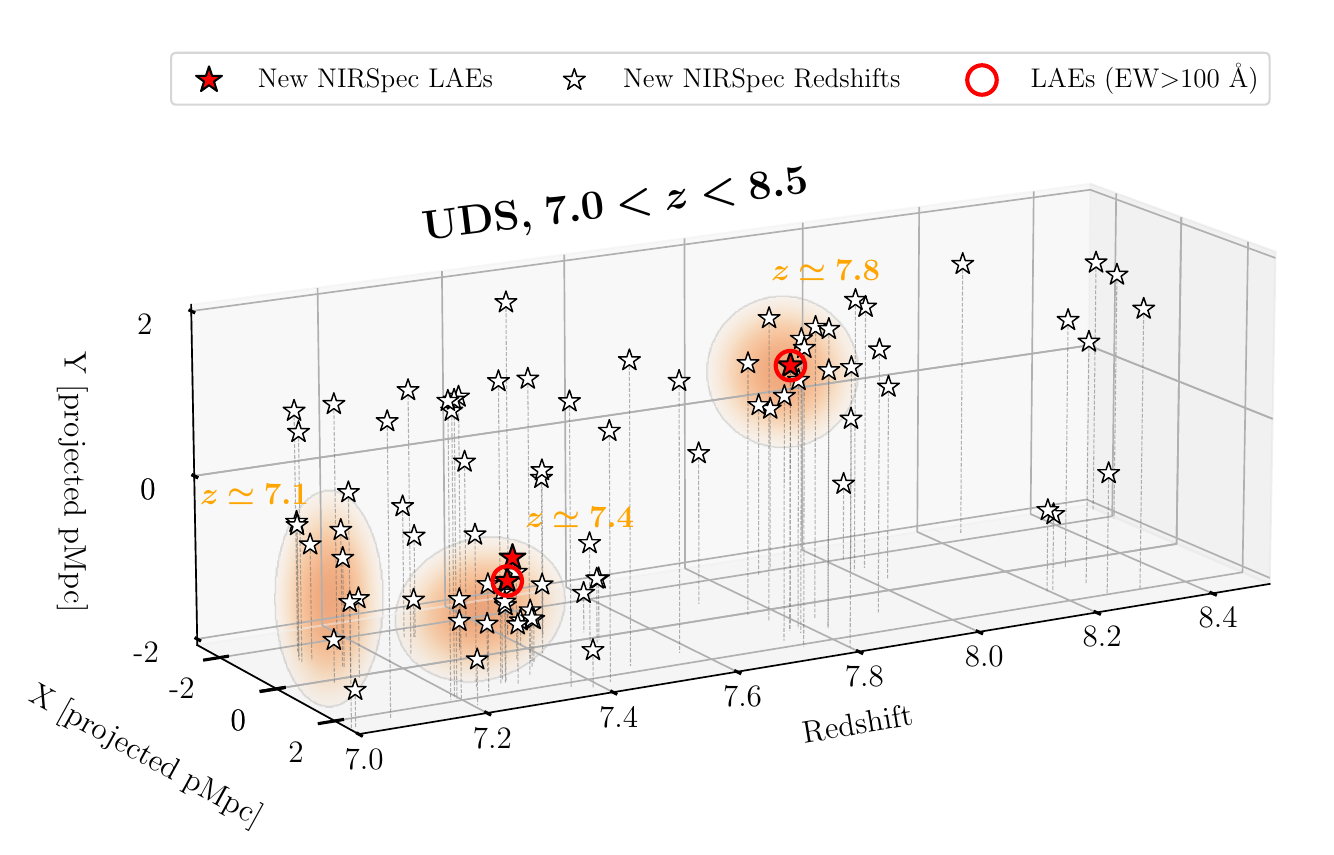} 
    \end{minipage}
    \begin{minipage}[c]{0.47\textwidth}
        \centering
        \includegraphics[width=\linewidth]{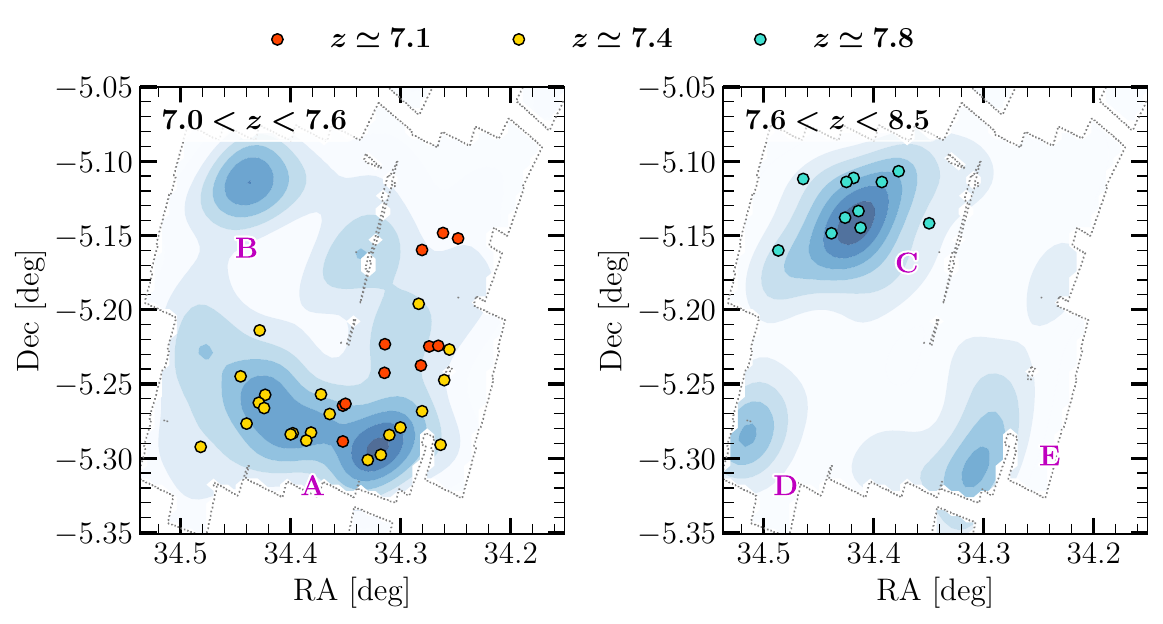}
    \end{minipage}
    \caption{Spatial distribution of NIRSpec-confirmed sources in the UDS field.
    The left panel shows the 3D distribution, with red stars corresponding to the newly-identified \lya{} emitting galaxies.
    Large red circles indicate galaxies with high \lya{} EWs ($>100$~\AA{}).
    We also plot  galaxies with redshift confirmation from NIRSpec but without \lya{} detections as open black symbols. Three overdensity candidates ($z\simeq 7.1$, 7.4, 7,8) are  illustrated with orange shaded colors.
    In the right two panels, we also show the 2D distribution of  NIRSpec-confirmed galaxies in the overdensities (colored dots) relative to the surface density (blue shaded colors) of photometrically selected sources at $z=7.0\mbox{--}7.6$ (middle panel) and $z=7.0\mbox{--}8.5$ (right panel).
    The  photometric overdensities are labeled with letters.}
    \label{fig:3dmaps_uds}
\end{figure*}

\begin{figure*}[t!]
    \centering
    \begin{minipage}[c]{0.52\textwidth}
        \includegraphics[width=\linewidth]{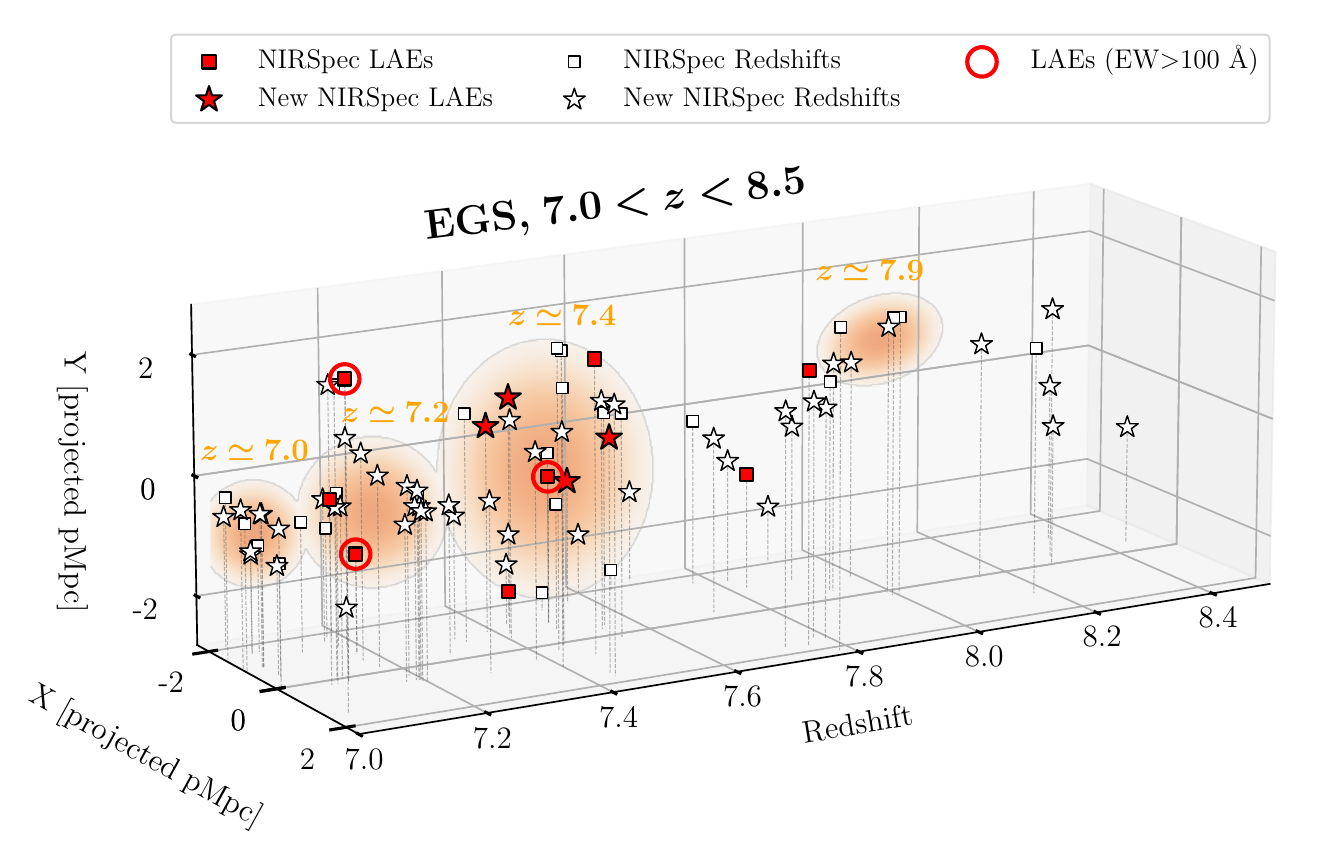} 
    \end{minipage}
    \begin{minipage}[c]{0.47\textwidth}
        \centering
        \includegraphics[width=\linewidth]{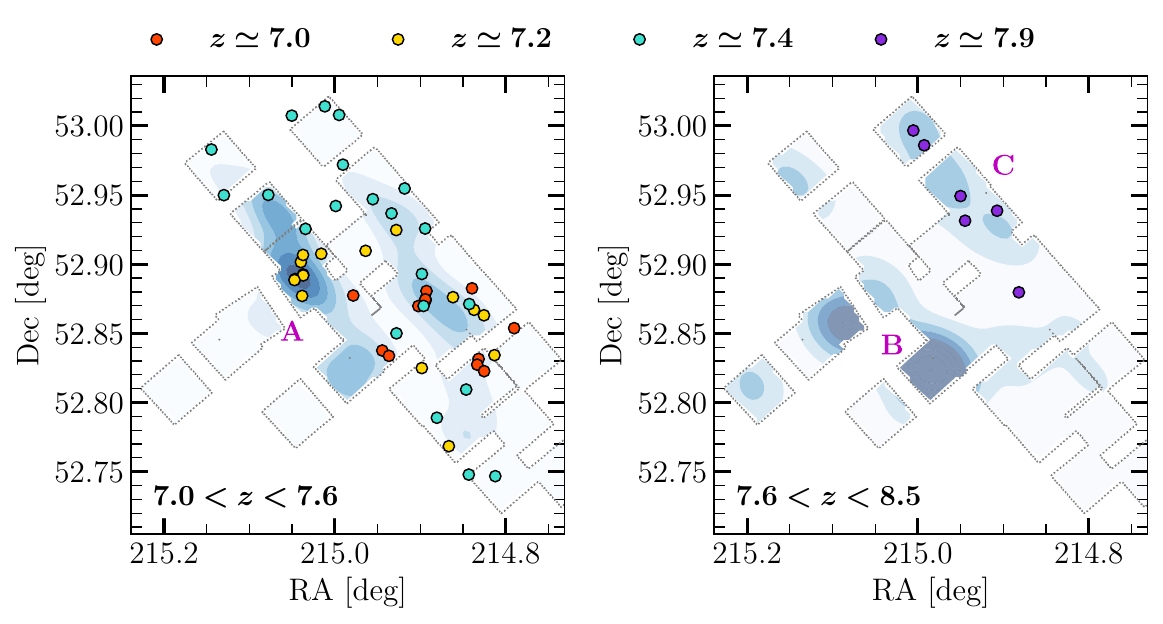}
    \end{minipage}
    \caption{Similar to Figure~\ref{fig:3dmaps_uds}, but for the spatial distribution of NIRSpec confirmed sources in the EGS field.
    The left panel shows the 3D distribution of the galaxies with redshift confirmation from NIRSpec.
    We show the \lya{} emitting galaxies with red filled symbols and galaxies without \lya{} detections with open symbols.
    Star symbols correspond to sources identified from the more recently-released observations during Cycles 2--3, and square symbols are those from earlier Cycle 1 NIRSpec spectra.
    The right two panels show how the spatial distribution of NIRSpec-confirmed overdensities compares to the photometric overdensities, considering separately sources at $z=7.0\mbox{--}7.6$ (middle panel) and $z=7.0\mbox{--}8.5$ (right panel).
    }
    \label{fig:3dmaps_egs}
\end{figure*}

\subsection{Identification of Overdensities in {\it JWST} Fields}\label{subsec:overden_identification}

The NIRCam grism provides the most reliable overdensity measurements, enabling the identification of emission line galaxies above a fixed flux threshold across an entire field. 
We follow the procedures described in \cite{Tang2024_nirspec} to identify $z\gtrsim 7$ galaxy overdensities in 62 arcmin$^2$ sub-regions of GOODS-N and GOODS-S using F444W grism observations from the First Reionization Epoch Spectroscopically Complete Observations (FRESCO; \citealt{Oesch2023}) program. At $z>7$, FRESCO identifies galaxy redshifts by detection of the  [\oiii{}]$\lambda\lambda4960,5008$ and H$\beta$ emission lines. We base our overdensity calculation on galaxies presented in the FRESCO team [\oiii{}] redshift catalog  \citep{Meyer2024}. Our methodology has been described in detail in \citet{Tang2024_nirspec}, but we 
briefly present the approach below for completeness. 

As a first step toward quantifying NIRCam grism overdensities, we measure the average number of galaxies over $z=7.0$ to 8.5. Here we consider only sources with robust emission line detections and redshift determinations. We adopt a fixed [\oiii{}]$\lambda5008$ flux limit of $2\times10^{-18}$ \ergscmA{} (corresponding to a typical $5\sigma$ detection), and we 
additionally require a quality flag of $q\geq2$ (as suggested by \citealt{Meyer2024}).  In each field, we measure the median number of galaxies per d$z=0.2$ bin (corresponding to a radial distance of $\simeq$6--8 pMpc)  using a large number (N=1000) of randomly chosen central redshifts between $z=7.0$ and $z=8.5$. 
We then search for redshift bins that appear overdense relative to the median.  We will define grism overdensities  as those redshift bins that appear 
$\geq3\times$ denser than the median. 
Our results for GOODS-S and GOODS-N are described in detail in \S~\ref{subsec:overden_goodss} and \S~\ref{subsec:overden_goodsn}, respectively.  

We note that while the grism does provide our most robust route to identifying overdense regions, there are several shortcomings.
Since the grism  selections at $z\gtrsim 7$ are limited to emission line objects, they will not select galaxies that are in an off mode of star formation (e.g., \citealt{Endsley2024_jades,Looser2024,Weibel2025}). Furthermore, at current sensitivities, grism-based measurements are mostly limited to UV luminous galaxies (M$_{\rm{UV}}<-\rm{19.7}$; \citealt{Meyer2024,Tang2024_nirspec}). If an overdense region does not have a 
large population of UV luminous systems, it is possible that it may not be selected in a grism survey.  

In fields where we do not have grism observations, we utilize the spatial distribution of NIRSpec-confirmed sources to identify overdense regions. 
Here we search for large-scale (mostly $\gtrsim 1$ pMpc) sub-regions in the fields that have galaxy densities in 
excess of that predicted from the UV luminosity function. We will describe the field-dependent spectroscopic overdensities in the following subsections.  However, we note that the spectroscopic  mapping of large scale structures may be incomplete due to the selection function and incomplete coverage. Our visual selection is meant to identify likely overdense regions for future spectroscopic follow-up, but we acknowledge that these selections may not include all structures in a given field.

To achieve a more systematic investigation in these fields, we also investigate the galaxy distribution utilizing the photometrically-selected samples. We focus this investigation on the UDS and EGS given the wide area imaging of both fields. We will discuss the photometric selections here, referring to them in more detail in the following subsections. We limit our analysis to the photometric sources brighter than the median 5$\sigma$ depth in each field (28.8 mag in EGS and 27.8 mag in UDS) to avoid low completeness.
For galaxies at $z=7.0\mbox{--}7.6$, the \oiiihb{} emission lines will be redshifted to 3.9--4.1~\um{}, resulting in a unique color excess in F410M relative to F444W filter (an EW \oiiihb{} of 400~\AA{} will lead to F410M - F444W = 0.60 mag).  This allows us to separate galaxies likely to lie at $z=7.0\mbox{--}7.6$ from those in $z=7.6\mbox{--}8.5$.

We present the angular distribution of galaxies in the UDS and EGS with photometric redshifts of $z=7.0\mbox{--}7.6$ and $z=7.6\mbox{--}8.5$ in Figure~\ref{fig:photoz}. We illustrate the galaxy surface density in blue, computed using a kernel density estimation approach where each source is represented by a Gaussian kernel, and the local density is computed as the sum of all kernel contributions at each position.  
To identify photometric overdensities, we place 1000 randomly-distributed circular apertures of radius $R = 2$ arcmin (corresponding to a projected scale of $\sim$0.6 pMpc at $z = 7.0\mbox{--}8.5$) across each field.
For each aperture, we compute the number of enclosed sources and normalize it by the effective area of the aperture that overlaps with the NIRCam footprint. 
We then calculate the median number of sources per aperture, considering only those apertures with at least 70\% areal coverage from NIRCam.
We locate the regions that are at least 2$\times$ more overdense than this median.
When multiple overlapping apertures are both considered overdense, we select that which has the highest amplitude as the center of the photometric overdensity. 

In total, the photometric method identifies 5 likely-overdense regions in the UDS, which are labeled alphabetically in Figure~\ref{fig:photoz}. We note that several of these may belong to the same overdense structure given the close redshift separation. In the EGS, the average surface density is twice that of the UDS (when adopting the shallower F150W magnitude limit of UDS), suggesting the total galaxy density is larger. We identify three peaks in the EGS galaxy distribution, one in each of the two redshift bins. The small number of individual overdensities in the EGS may suggest that the entire footprint traces a large ($>$4.5 pMpc) overdensity. We will discuss this possibility in the following sections. 

In what follows, we provide an overview of the distribution of galaxies in the five fields considered in this paper, identifying spectroscopic and photometric overdensities. We then briefly detail the position of Ly$\alpha$ emitters relative to these structures. The goal of these subsections is to characterize the current state of observations in these fields. In \S5, we discuss implications for the likely sizes of ionized regions at $z\simeq 7.0-8.5$.
We will summarize the spectroscopic overdensities in Table~\ref{tab:overdensity_summary}.

\subsection{Overdensities and Ly$\alpha$ in the UDS Field}\label{subsec:overden_uds}

In the last year, prism observations have provided our first look at the distribution of $z\gtrsim 7$ galaxies in the UDS field. Our redshift catalog includes a total of 84 galaxies at $7.0<z<8.5$ distributed over 151 arcmin$^2$, with 67 galaxies from RUBIES and 17 galaxies from CAPERS.
The three-dimensional map of galaxies with confirmed redshifts is shown in Figure~\ref{fig:3dmaps_uds}. We first provide an overview of candidate structures in the map before discussing each in more detail below.  The spatial distribution demonstrates that the majority of galaxies are at $z\simeq 7.3-7.4$, consistent with the distribution of redshifts in the field (Figure~\ref{fig:zhist_nirspec}). The redshift histogram also shows a minor peak at $z\simeq 7.8$, which appears as a clustered group of galaxies in the spatial map. There is an additional network of galaxies at $z\simeq 7.1$, which we also highlight as a potential overdensity. While these identifications are visual, we will quantify the spectroscopic overdensities in these regions below, and we will investigate whether they overlap with the known photometric overdensities.

The potential large scale structure at $z\simeq 7.1$ includes 11 galaxies with confirmed redshifts between $z=7.08$ and $z=7.16$. The redshift spread of the galaxies corresponds to a radial length of 2.1 pMpc. As is clear in Figure~\ref{fig:3dmaps_uds}, the galaxies at this redshift are not tightly clustered, spanning  10.3 arcmin across the field, equivalent to a projected distance of 3.2 pMpc. The 11 galaxies can be fit in a rectangular area of 10.3$\times$2.9 arcmin$^2$. 
We show the angular distribution of $z\simeq 7.1$ galaxies (red circles) compared to the surface density of photometric candidates (blue contours) in the middle panel of Figure~\ref{fig:3dmaps_uds}. We see that the galaxies are spread across several of the peaks in the photometric distribution of sources, perhaps contributing somewhat to the photometric structure denoted A. We finally consider whether the existing data show sufficient evidence for a spectroscopic overdensity at $z\simeq 7.1$.
If we adopt the luminosity function of \cite{Bouwens2021_uvlf}, we would predict 4.3$\pm$0.2 galaxies with $M_{\rm UV}<-19$ in the volume sampling this area from $z=7.08$ and $z=7.13$.  Given that 7 of the 11 galaxies in this area are above this luminosity threshold, we find spectroscopic evidence for a mild (1.6$\times$)  overdensity. This value is a lower limit, as existing spectroscopic samples are incomplete.

The  structure we identify at $z\simeq 7.4$ can be divided into two substructures. The first of which contains 13 galaxies  at $z=7.24-7.32$ (d$z$=0.08; 3 pMpc) spanning an angular diameter of 11.3 arcmin (3.5 pMpc). The source density peaks near the apparent center of the structure with 9 galaxies confirmed at $z=7.28-7.32$, a radial length of 1.4 pMpc, spanning a rectangular area of 10.8$\times$2.8 arcmin$^2$ (3.1 pMpc $\times$ 0.9 pMpc). We expect 1.7$\pm$0.1 galaxies with $M_{\rm UV}<-19$  within this volume from the \cite{Bouwens2021_uvlf} luminosity function, indicating the central region is likely significantly ($4.7\times$) overdense. The second substructure includes 8 $M_{\rm UV}<-19$ galaxies at $z=7.36-7.43$ (a radial length of 2.7 pMpc) spanning an angular diameter of 7.0 arcmin (2.1 pMpc).  When compared to expectations from the luminosity function, we find evidence that this substructure is at least 2.2$\times$ overdense. Two extremely luminous galaxies ($M_{\rm UV}\approx-21.5$) are found among the 8 confirmed galaxies. 
Both of the $z\simeq 7.3-7.4$ substructures appear clustered on the photometric overdensity A shown in the upper left panel of Figure~\ref{fig:photoz} (see also yellow circles in the middle panel of Figure~\ref{fig:3dmaps_uds}). Hence there is evidence for both photometric and spectroscopic overdensities in the UDS at $z\simeq 7.3-7.4$.

The highest redshift association of galaxies we consider in the UDS is comprised of 11 sources at $z=7.74-7.83$, spanning a radial distance of 3.2 pMpc. The galaxies are 
distributed over an angular scale of 8.3 arcmin (2.4 pMpc).  The rectangular 
area covered by the galaxies (8.3$\times$2.7 arcmin$^2$) should contain 
1.9$\pm$0.1 galaxies with $M_{\rm UV}<19$  over the $d$z$=0.09$ redshift window according to the \cite{Bouwens2021_uvlf} UV luminosity function. The structure consists of 8 galaxies that meet this luminosity threshold, suggesting an overdensity that is at least 4.2$\times$ the average. 
The spatial distribution of targets in this  structure appears to overlap with the photometric overdensity C (see the right panel in Figure~\ref{fig:3dmaps_uds}).

Our discussion  has thus far  focused on large-scale overdense regions, but we note that there is a potential smaller-scale structure at $z\simeq 7.2$. Four closely situated galaxies (0.22~pMpc in projection) are found at  $z=7.18 - 7.20$, spanning a radial distance of 0.63 pMpc (Figure~\ref{fig:3dmaps_uds}). The small size is similar to the strong  overdensity previously reported at $z\simeq 7.89$ in Abell 2744 (\citealt{Morishita2023}, \S~\ref{subsec:overden_a2744}) and may reflect a compact  group of galaxies.

There are three photometric overdensities that do not appear to show spectroscopic counterparts (B, D, E; Figure \ref{fig:photoz} and \ref{fig:3dmaps_uds}). 
At $7.0<z<7.6$, Region B contains 19 galaxies within a $R=2$ arcmin aperture, which, when compared to the average number (7.7), implies a 2.6$\times$ photometric overdensity (after taking into account the fraction of the aperture not covered with imaging).
Both regions D and E are at $7.6<z<8.5$, each containing 8--9 sources within a $R=2$ arcmin aperture.
Compared to the field average (3 per $R=2$ arcmin aperture), both numbers correspond to photometric overdensities at $\approx3.2\times$ after taking into account the area within the aperture not covered by imaging.
As future spectroscopic efforts target more galaxies in the UDS, we may expect to find evidence for structures associated with these regions.

The CAPERS and RUBIES prism spectra provide shallow \lya{} emission constraints for 69 of the $z\simeq 7.01-8.50$ galaxies described above. Only three galaxies in this sample have been found with  \lya{} emission (Table~\ref{tab:sample}), all of which 
appear associated with likely overdense regions (Figure~\ref{fig:3dmaps_uds}).
One of the large Ly$\alpha$ EW galaxies, RUBIES-UDS-24303 (EW = $200_{-78}^{+50}$~\AA{}) is part of the structure at $z\sim7.3$. 
The other strong \lya{} emitter CAPERS-UDS-142615 (EW = $284_{-75}^{+56}$~\AA{}) lies near the center of the candidate overdense region at $z\sim7.8$.
We note that other galaxies in these two regions have only shallow upper limits on the \lya{} EW (median 78~\AA{}), so deeper spectra could reveal moderate strength Ly$\alpha$ in many.  The presence of such 
intense Ly$\alpha$ emission in the two overdense spectroscopic structures 
may already point to an enhanced transmission of Ly$\alpha$, as 
would be expected in large ionized sightlines.  We will describe this in more detail in \S~\ref{sec:results_lya}.

\subsection{Overdensities  and Ly$\alpha$ in the EGS Field}\label{subsec:overden_egs}

Our spectroscopic sample includes 89 galaxies with NIRSpec-based redshifts in the EGS field at $z=7.0\mbox{--}8.5$. While the CEERS ERS observations contributed many of these measurements, more recent surveys are making an increasingly important 
contribution, with 32 sources from RUBIES, 10 from GO 4287, and 13 from CAPERS (see \S2). The current area sampled by spectroscopy in the EGS is  129 arcmin$^2$, and the total spectroscopic sample size is nearly three times greater than that which was reported after the CEERS observations in Cycle 1, allowing a much-improved map of the spatial distribution of galaxies in the EGS (Figure~\ref{fig:3dmaps_egs}).

The overdensities in the EGS have been the subject of several papers over the last decade, both with \hst{} imaging \citep{Leonova2022} and early \jwst{} Ly$\alpha$ spectroscopy \citep{Tang2023,Chen2024,Napolitano2024,Tang2024_nirspec}. While NIRSpec 
observations are not complete, we are able to identify candidate overdensities. The redshift histogram reveals several peaks (Figure~\ref{fig:zhist_nirspec}), the strongest of which are at $z\simeq 7.0-7.2$, $z\simeq 7.5$, and $z\simeq 7.9$. The NIRSpec map shown in the left panel of Figure~\ref{fig:3dmaps_egs} shows these peaks correspond to potential galaxy structures at $z\simeq 7.0$, $z\simeq 7.2$, $z\simeq 7.5$, and 
$z\simeq 7.9$. We will discuss each of these in more detail below, quantifying evidence that the galaxies at these redshifts present spectroscopic overdensities.

The  $z\simeq 7.0$ structure consists of 13 galaxies at $z=7.00-7.06$, spanning a radial distance of 2.6 pMpc. The confirmed systems appear clustered in a region that is 
7.0$\times$3.5 arcmin$^2$ in area. The \citet{Bouwens2021_uvlf} UV luminosity function predicts that we should find 2.5$\pm 0.1$ galaxies brighter than 
M$_{\rm{UV}}=-19.0$ in an area of this size spanning from $z=7.00$ to $z=7.60$. Of the 13 galaxies that are confirmed in this structure, 8 galaxies appear brighter than this threshold, suggesting an overdensity with an amplitude of at least 3.2$\times$. We note that there are an additional 28 galaxies from $z=6.93$ to $z=7.00$ in our redshift catalog (see Table~\ref{tab:egs_z6p9}), so it is conceivable this structure extends slightly below the 
the lower bound of our redshift cut.

The $z\simeq 7.2$ structure is comprised of 16 galaxies at $z=7.16\mbox{--}7.20$, a radial distance of 1.6 pMpc. The galaxies are strongly clustered in one of the strongest photometric overdensities (region A, see Figure~\ref{fig:3dmaps_egs} middle panel), but they also extend across a larger fraction of the footprint, spanning an angular scale of 10.3 arcmin (3.2 pMpc). If we consider the rectangular area that covers the galaxies (10.3$\times$5.0 arcmin$^2$), we would expect to 
recover 3.1$\pm$0.2 galaxies brighter than M$_{\rm{UV}}=-19.0$ based on the 
luminosity function of \citet{Bouwens2021_uvlf}. The existing catalog reveals 11 galaxies with luminosities above this threshold, implying an overdensity that is at least 3.5$\times$ average.

The $z\simeq 7.4$ association of galaxies consists of 22 systems with 
confirmed redshifts at  $z=7.38-7.56$, suggesting the potential presence of a structure spanning a radial distance of 6.0 pMpc. Unlike the structures described above, the galaxies at $z\simeq 7.4-7.6$ are spread throughout most of the EGS footprint, with an angular scale of 18.6 arcmin (5.6 pMpc). Over this larger volume, the existing spectroscopy does not indicate an overdensity. However, there are several potential substructures.
The first structure contains 6 galaxies with redshift at $z=7.43\mbox{--}7.47$, extending over a line of sight distance of 1.4~pMpc and a projected distance of 1.8~pMpc.
This region also host a very luminous galaxy, CEERS-698 ($M_{\rm UV}=-21.7$), which was confirmed prior to \jwst{} \citep{Roberts-Borsani2016,Stark2017}.
The \cite{Bouwens2021_uvlf} luminosity function predicts an average of $0.7_{-0.1}^{+0.1}$ galaxies over the rectangular area (5.6$\times$2.6~arcmin$^{2}$) occupied by these galaxies.
Four (of the six) galaxies are brighter than this magnitude limit, suggesting it to be greater than 5.4$\times$ overdense.
The second structure includes another 6 galaxies over a comparably narrow redshift range of $z=7.45\mbox{--}7.49$ (a radial distance of 1.3~pMpc) but with a slightly larger angular scale of 9.1~arcmin (2.8~pMpc projected distance).
By comparing to the prediction of the UVLF, we also find this substructure to be at least $2.8\times$ overdense. The two structures are separated by roughly 1.8 pMpc. More extensive spectroscopy over the field is required to explore whether the two $z\simeq 7.4$ substructures are connected.

Previous studies have described an association of galaxies in the EGS at $z\simeq 7.7$, with spectroscopically confirmed galaxies and a known 
Ly$\alpha$ emitter from Keck spectroscopy \citep{Oesch2015}. The current 
data do not identify the $z\simeq 7.7$ galaxies as overdense, perhaps 
simply a result of spectroscopic incompleteness. However, there is 
a candidate structure at $z\simeq 7.9$, with 
six galaxies spread across $z=7.90-7.99$ and spanning a radial length of 2.9 pMpc. The galaxies are situated over a rectangular area of 8.3$\times$1.4~arcmin$^2$(2.4$\times$0.4~pMpc$^2$), along the upper 
region of the EGS footprint, where photometry indicates a potential 
overdensity (C, see right panel of Figure~\ref{fig:3dmaps_egs}).
Over this area, the luminosity function predicts $0.9_{-0.1}^{+0.1}$ galaxies with $M_{\rm UV}<-19$ while we find 3, suggesting it to be at least a 3.3$\times$ overdensity.

The NIRSpec database currently provides constraints on  \lya{} emission in 71 galaxies from $z=7.00\mbox{--}8.44$, a significant improvement from earlier studies in the EGS.
We detect \lya{} emission in 12  galaxies, 10 of which are found within the redshift range of $z=7.10$--7.56, corresponding to a radial sightline of 18 pMpc. Three of the 
\lya{} emitters are extremely strong (EW$>$100~\AA{}), each of 
which lies at redshift associated with an overdensity (the $z\simeq7.2$ or $z\simeq7.4$ structure). One of the strongest Ly$\alpha$ emitters (CEERS-44 at $z=7.10$) appears spatially offset (3.3 pMpc) from the region we have identified as spectroscopically overdense at $z\simeq 7.2$ (Figure~\ref{fig:3dmaps_egs}), but this is plausibly just due to incompleteness in the spectroscopic coverage across the field. We will discuss the Ly$\alpha$ detections in EGS in more detail in \S~\ref{subsec:egs_ionized}.

\subsection{Overdensities  and Ly$\alpha$ in the GOODS-S Field}\label{subsec:overden_goodss}

\begin{figure}
    \centering
    \includegraphics[width=1\columnwidth]{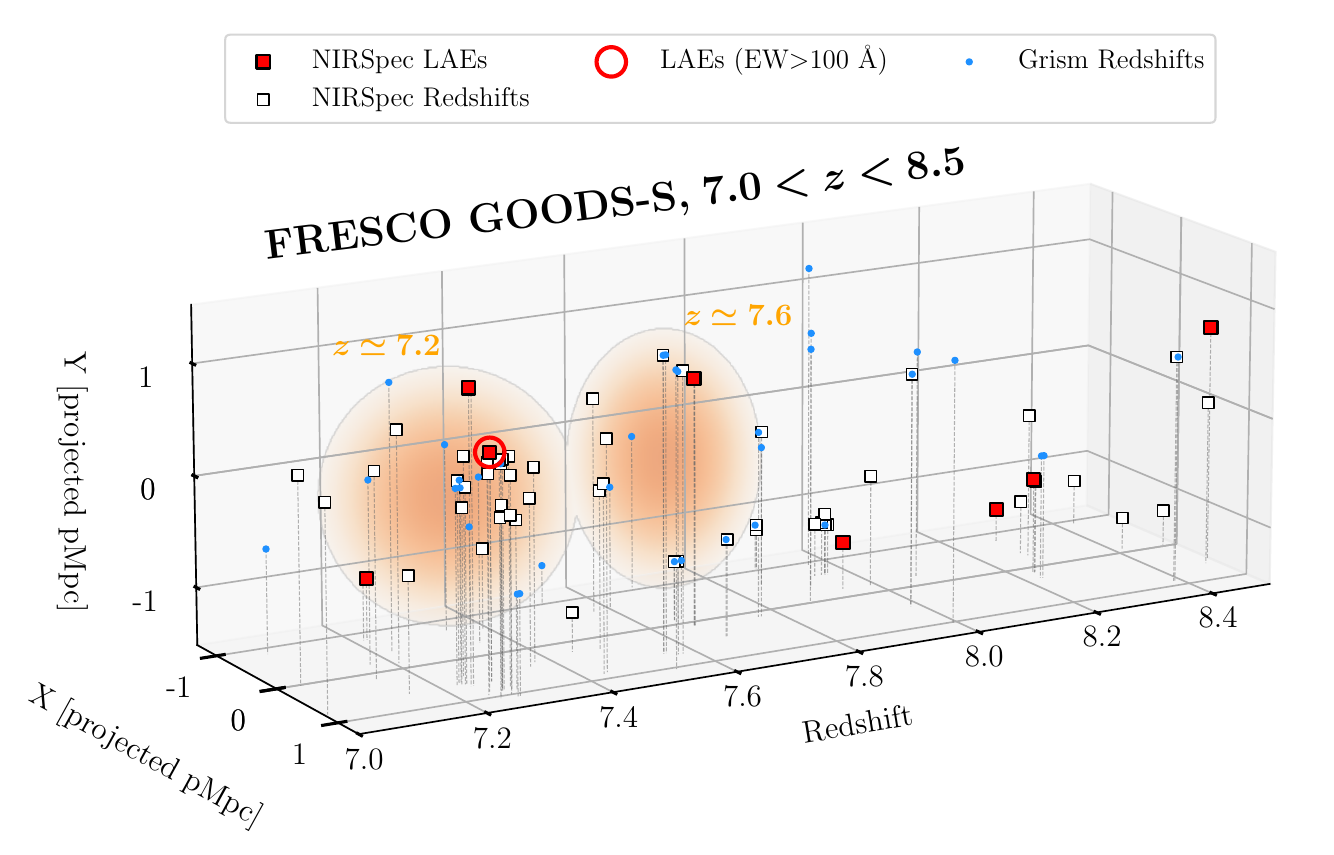}
    \includegraphics[width=1\columnwidth]{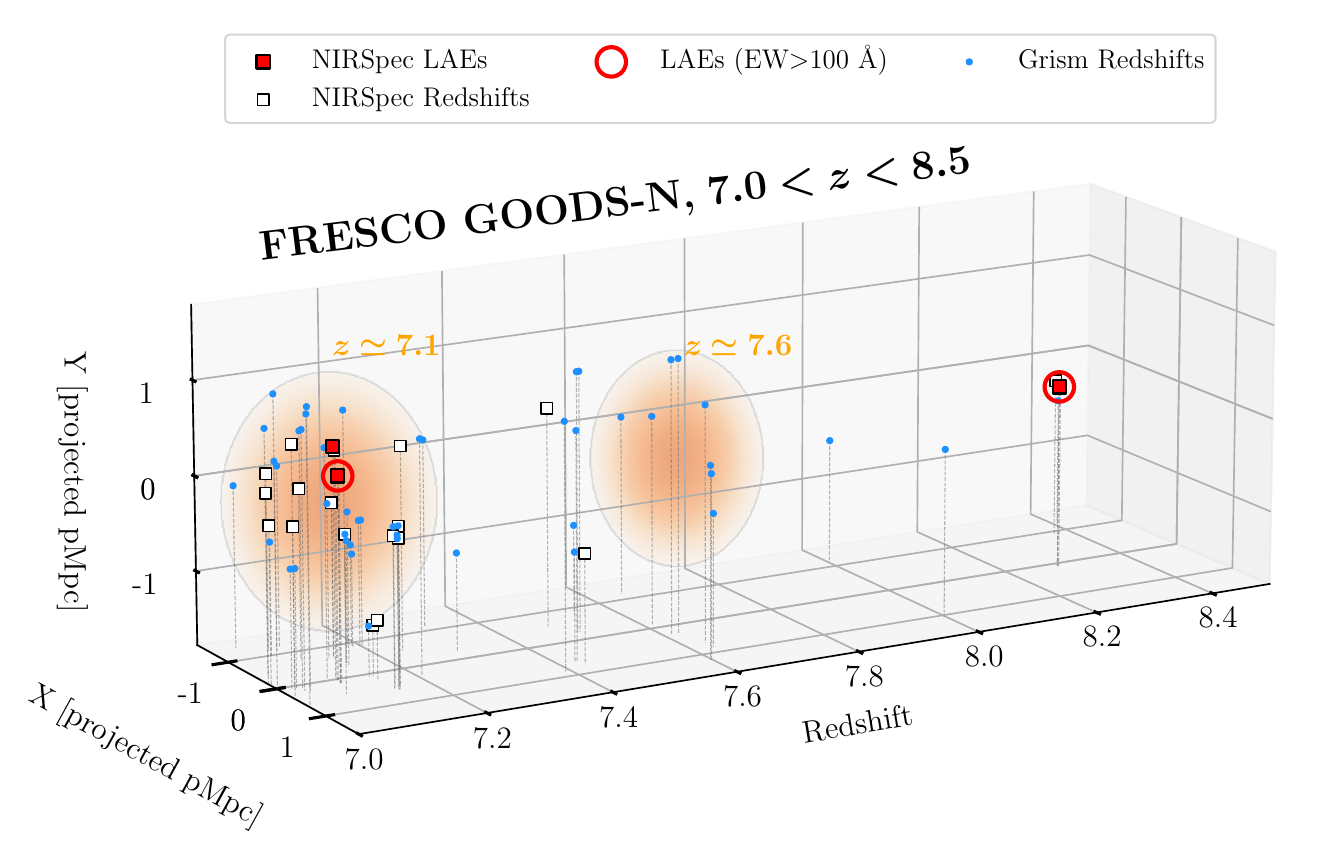}
    \includegraphics[width=1\columnwidth]{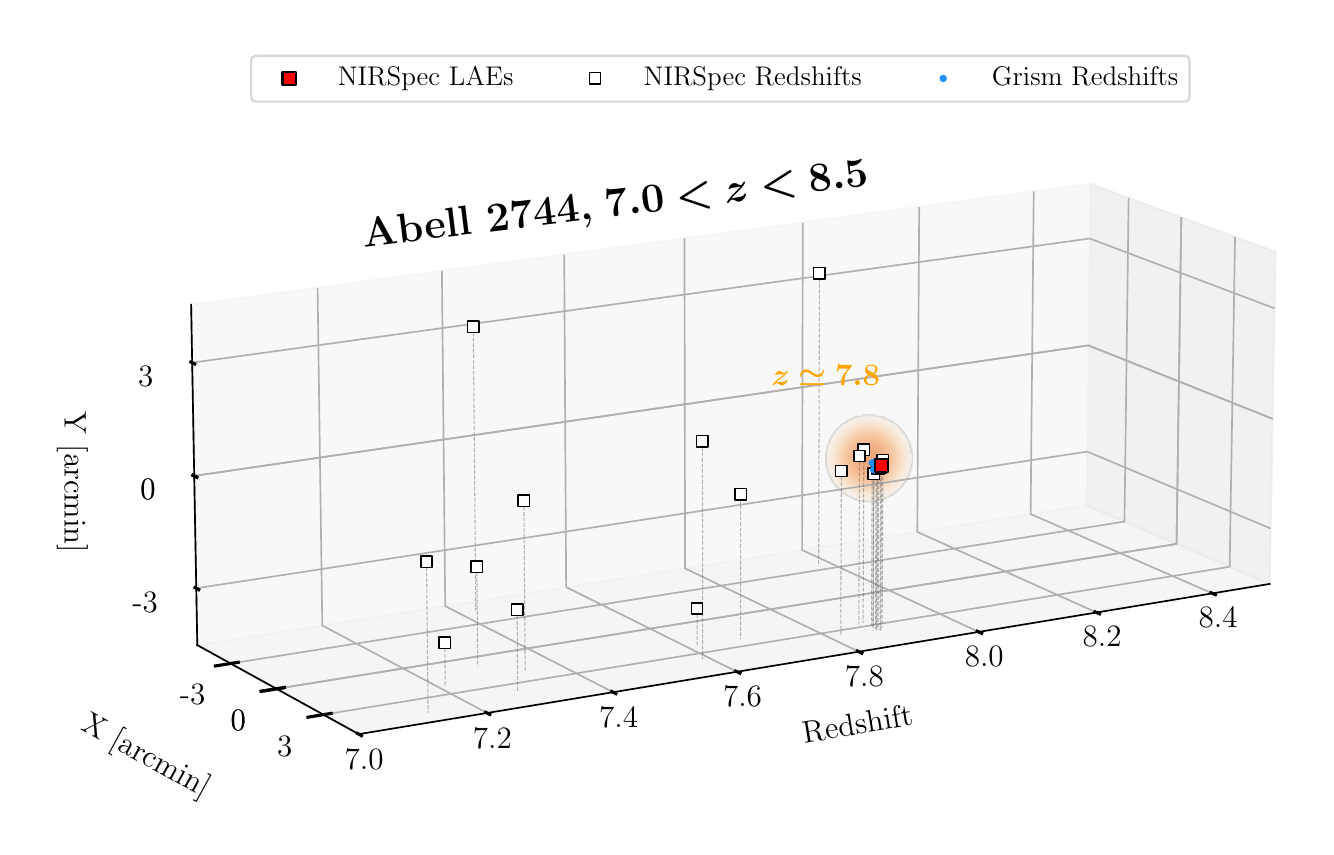}
    \caption{The 3D  distribution of the NIRSpec-confirmed galaxies relative to overdensities in the GOODS-S (top), GOODS-N (middle), and Abell 2744 (bottom) fields.
    Here, in the GOODS-S and GOODS-N fields, we only consider the NIRSpec sample that is within the footprint of FRESCO  NIRCam grism observations.
    Red squares are \lya{} emitting galaxies, while large red circles indicate those with the largest Ly$\alpha$ EWs ($>100$~\AA{}).
    We also plot the galaxies with NIRSpec confirmation but without \lya{} detections as open black symbols (stars if from GTO 1286+1287 in GOODS-S, and squares if from earlier observations).
    We characterize the environment with available NIRCam grism observations from FRESCO in each field (blue dots), with the identified overdense regions and their redshifts shown in light orange.}
    \label{fig:3dmaps_goods_a2744}
\end{figure}

\begin{figure*}
    \centering
    \includegraphics[width=0.42\textwidth]{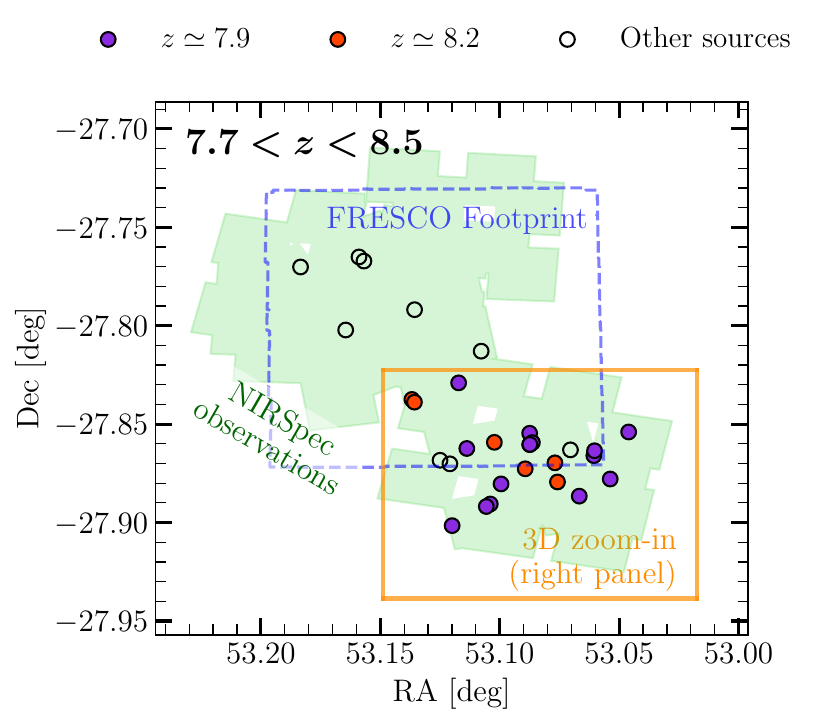}
    \includegraphics[width=0.52\textwidth]{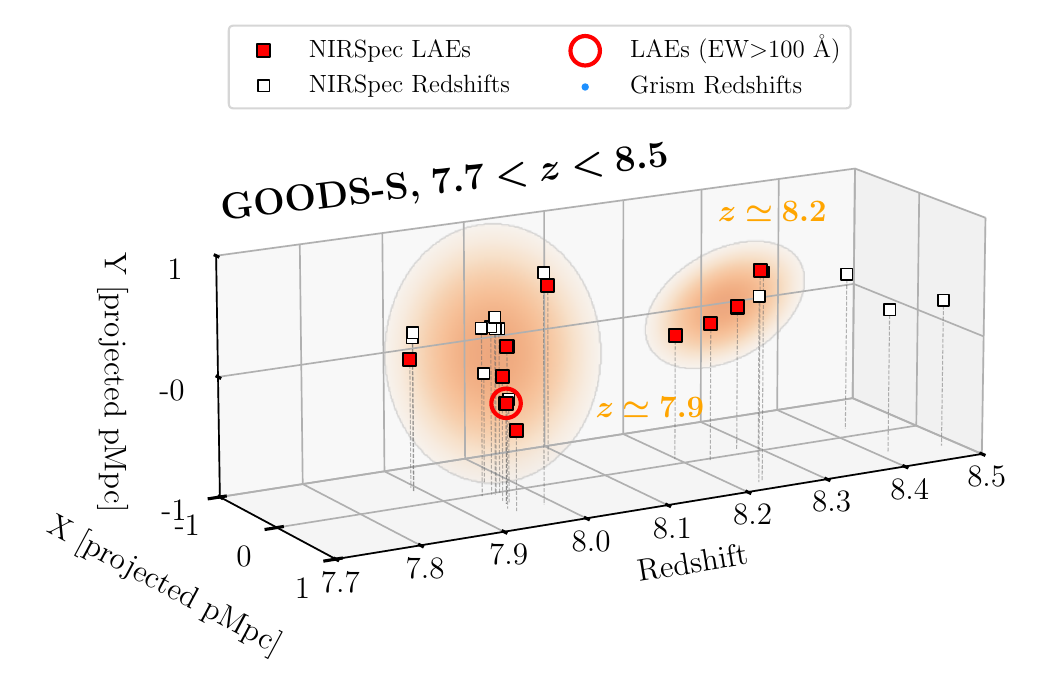}
    \caption{
    (Left:) The 2D-distribution of NIRSpec-confirmed galaxies at $7.7<z<8.5$ across GOODS-S. Several potential overdensity candidates are located outside of the FRESCO footprint in a region denoted with an orange square. The green shading illustrates the region covered with NIRSpec observations. (Right:) Three-dimensional distribution of galaxies  in the region highlighted by the orange square in the left panel. Two potential overdense structures are seen  ($z\simeq 7.9$ and $z\simeq 8.2$), neither of which is identified in the grism observations given their location outside of the FRESCO footprint. 
    }
    \label{fig:goodss_z8}
\end{figure*}

The spatial distribution of $z>7$ [\oiii{}] emitters in the FRESCO footprint of GOODS-S has been quantified in several papers \citep{Helton2024,Tang2024_nirspec}, as has the connection of Ly$\alpha$ emitters and overdensities \citep{Tang2024_nirspec}. 
Here we update earlier investigations to include new NIRSpec observations in GOODS-South from the JADES team \citep{D'Eugenio2024}. In particular, the 1286 and 1287 programs of JADES add  45 new spectroscopically confirmed galaxies at $z=7.0\mbox{--}8.5$, resulting in a total NIRSpec sample of 75 galaxies in this redshift range.

We briefly describe the overdensities identified in the FRESCO footprint of GOODS-S, though our analysis largely follows what we have presented in \citet{Tang2024_z56}. 
Two peaks are seen in the 
redshift distribution of [\oiii{}] emitters, one at $z\simeq 7.2$ (8 galaxies over $z=7.16-7.36$) and the other at $z\simeq 7.6$ (9 galaxies over $z=7.52-7.72$). This translates into an overdensity of 4$\times$ ($z\simeq 7.2$) and 4.5$\times$ ($z\simeq 7.6$). We note that the overdensity at $z\simeq 7.6$ would have a slightly higher amplitude if we were to account for the redshift-dependence of the [\oiii{}] emitter number density. However, the precise effect is difficult to quantify given that the current [\oiii{}] luminosity functions are based on a small number of fields with overdensities contributing to the redshift-dependent evolution. In this paper, the amplitude of the overdensity is not critical to our analysis, so we will proceed with the measurements quoted above.

The JADES NIRSpec observations have taken the first steps to characterize the distribution of Ly$\alpha$ emitters in GOODS-S (Figure~\ref{fig:3dmaps_goods_a2744} top panel, see previous analyses in \citealt{Tang2024_nirspec,Witstok2024}). We first limit our discussion to those galaxies observed within the FRESCO footprint (see Table~\ref{tab:sample}). Our catalog 
includes 18 galaxies with Ly$\alpha$ constraints in the $z\simeq 7.2$ ($z=7.20-7.29$) overdensity, with typical 3$\sigma$ EW limits of 41~\AA. This sample includes three Ly$\alpha$ emitters, one of which has extremely strong Ly$\alpha$ emission (EW=$244_{-27}^{+21}$~\AA) first reported in \cite{Saxena2023}. 
Meanwhile, the $z\simeq 7.6$ overdensity contains 7 galaxies with Ly$\alpha$ constraints within the FRESCO footprint. One shows moderate strength (EW=$24_{-2}^{+2}$~\AA) Ly$\alpha$ emission 
(see also \citealt{Tang2024_nirspec}), while the others reveal non-detections with typical 
EW limits of 30~\AA. Larger samples will be required to verify whether these volumes enable enhanced transmission, as we will discuss in \S~\ref{sec:results_lya}.

Within the FRESCO footprint, there are also four Ly$\alpha$ detections (all at $z>7.8$) not associated with the grism overdensities mentioned above. This may appear 
surprising if the escape of Ly$\alpha$ is linked to the enhanced transmission associated with overdensities. 
However, these four Ly$\alpha$ detections appear situated near the 
edge of the FRESCO footprint. It is conceivable that these trace 
an overdense structure that extends off of the footprint. Indeed, we find that three of the \lya{} emitters near the edge (those at $z\simeq$7.9--8.3) are surrounded by 4--6 neighboring galaxies (offset from the FRESCO footprint) with similar NIRSpec redshifts (d$z<0.1$) and close separations ($\sim3.5$~arcmin).

To investigate this further, we show the two-dimensional distribution of the sources confirmed with NIRSpec at $7.7<z<8.5$, now also considering those outside the grism footprint (Figure~\ref{fig:goodss_z8}). We find that the majority of the galaxies within the redshift range are contained within two structures.
The first one at $z\simeq7.9$ contains 16 galaxies spanning $z=7.89\mbox{--}8.07$ (line of sight distance 5.9~pMpc) over a 4.4~arcmin angular scale (1.3~pMpc). Of the 16 galaxies, 11  are found within a narrow window of $z\sim7.94\mbox{--}7.97$ (d$z=0.03$), including one UV luminous galaxy (\muv{}=$-21.1$).  Comparison to the \cite{Bouwens2021_uvlf} luminosity function indicates a spectroscopic overdensity factor of $\sim$5.1. Considering the full JADES spectroscopic dataset, the $z\simeq 7.9$ overdense structure hosts  6 \lya{} detections, including one high-EW \lya{} galaxy located outside the grism footprint: JADES-GS-12326 (EW = $120_{-9}^{+7}$~\AA{}; also see \citealt{Jones2025}).
The second galaxy structure that we find offset from FRESCO, at $z\sim8.2$, consists of 6 galaxies spanning $z=8.20-8.28$ (line of sight distance 2.3~pMpc) and an angular scale of 4.1~arcmin (1.2~pMpc).
All 6 galaxies are brighter than $M_{\rm UV}=-19$, which, when compared to the prediction from UVLF over this volume ($N = 0.1_{-0.1}^{+0.1}$ within $4.1\times0.5$~arcmin$^2$), implies this region to be a strong overdensity with a factor of at least 60 times.
We note that the overdensity factors will be lower if we adopt a larger area (7.4$\times$ when adopting a minimum rectangle width of 1~pMpc, i.e., over an rectangular area of $4.1\times3.5$~arcmin$^2$).
Similarly, 4 of the 6 galaxies at $z\sim8.2$ show \lya{} detections, with EW estimated to range from 23--28~\AA{} (median 26~\AA{}).
These results indicate that there are significant galaxy structures located at the redshifts of the $z>7.8$ Ly$\alpha$ emitters in the GOODS-S FRESCO footprint, despite there being  no strong evidence  for  overdensities  in FRESCO. Some caution needs to be taken when interpreting seemingly-isolated \lya{} emitters (particularly those near the edge of grism surveys), as they may still be part of structures primarily located off of the observational footprint.

\subsection{Overdensities  and Ly$\alpha$ in the GOODS-N Field}\label{subsec:overden_goodsn}

Several previous studies have investigated the distribution of [\oiii{}] emitters at $z>7$ in GOODS-N from the FRESCO dataset \citep[e.g.,][]{Helton2024,Meyer2024,Tang2024_nirspec}. Our results in this field are very similar to those presented in \citet{Tang2024_nirspec} as our sample does not include any additional spectroscopy not included in that paper. 
We briefly summarize the overdensities and \lya{} observations below.

As described in \S~\ref{subsec:overden_identification}, we have used the redshift distribution of the FRESCO [\oiii{}] emitters to isolate overdense regions.
We identify two redshifts with a significant excess of [\oiii{}] emitters: one is at $z\simeq7.1$, and the other is at 7.6 (see the middle panel of Figure~\ref{fig:3dmaps_goods_a2744}).
The first region hosts 22 [\oiii{}] emitters in the narrow redshift window of $z=7.00$–7.20 (8.4 pMpc along the line of sight), which is an $11\times$ enhancement relative to the average.
The second region includes 9 galaxies detected at $z=7.48$--7.68 (line of sight 7.2 pMpc), corresponding to a $4.5\times$ overdensity.
Both structures have been identified in previous studies, with overdensity estimates broadly consistent with what is reported here \citep{Helton2024, Meyer2024, Tang2024_nirspec}.

The JADES NIRSpec observations (prism and G140M) provide \lya{} constraints for 26 galaxies spanning $z=7.00\mbox{--}8.37$, which are also shown in the middle panel of Figure~\ref{fig:3dmaps_goods_a2744}.
We detect \lya{} emission in three galaxies, including two with very large EWs ($>100$~\AA{}), all previously reported \citep[e.g.,][]{Tang2024_nirspec,Witstok2024,Jones2025,Kageura2025}.
The spatial distribution of these \lya{} emitters with respect to the grism overdensities has been discussed before \citep{Tang2024_nirspec,Witstok2025_z8}, and our findings are similar.
Two out of the three LAEs, including one with large EW (JADES-GN-13041, $134_{-13}^{+9}$~\AA{}), are found associated with the overdensity at $z\simeq 7.1$. This redshift is also host to another ground-based LAE discovered over a decade ago \citep{Ono2012}.
The third, JADES-GN-1899 (EW = $118_{-12}^{+10}$\AA{}), lies at $z=8.28$ with no clear evidence of an accompanying overdensity (as noted in \citealt{Tang2024_nirspec}). It is thought that this system may have its Ly$\alpha$ transmission enhanced by its hard radiation field, enabling relatively strong Ly$\alpha$ emission in a small ionized bubble. 
Among the grism sources confirmed to be in overdensities, the NIRSpec observations provide \lya{} constraints for six galaxies, all part of the $z\simeq7.1$ structure. Two of the six galaxies reveal \lya{} detections, with one (JADES-GN-13041) having high EW.

\subsection{Overdensities  and Ly$\alpha$ in the Abell 2744 Field}\label{subsec:overden_a2744}

Our NIRSpec sample in the Abell 2744 field includes 19 galaxies spanning $z=7.13\mbox{--}7.98$ across $\approx$47 arcmin$^{2}$.
The majority (12/19) of them are identified from the GLASS ERS, UNCOVER, and DDT 2756 observations taken in Cycle 1. Observations from Cycle 2 confirmed an additional 7 galaxies, with 5 from a new UNCOVER pointing and 2 from program GO 3073 (see Table~\ref{tab:sample}). 
The spatial distribution of these galaxies is shown in the bottom panel of Figure~\ref{fig:3dmaps_goods_a2744}.

This field is known to host a significant overdensity at $z=7.88$ \citep{Hashimoto2023,Morishita2023}, as is clear looking at 
the redshift histogram in Figure~\ref{fig:zhist_nirspec}. The current spectroscopic sample 
includes 8 galaxies in this structure, with redshifts spanning $z=7.88\mbox{--}7.89$, all reported previously in the literature \citep{Hashimoto2023,Morishita2023,Chen2024}.
These 8 galaxies lie within a radius of 12 arcsec, corresponding to a projected radius of 60 pkpc (after correcting for lensing magnification; \citealt{Morishita2023}).
Based on the \cite{Bouwens2021_uvlf} UV luminosity function, we
expect no galaxies with $M_{\rm UV}<-19$ in such a small volume.
In contrast, 7 of the 8 galaxies are brighter than this magnitude threshold, suggesting this small area is significantly overdense
($\sim3500\times$), consistent with what was found by \cite{Morishita2023}.

NIRSpec spectra allow us to constrain \lya{} emission in all 19 galaxies in our Abell 2744 sample.  Initial work based on 7 $z\simeq 7.9$ spectra taken by the GLASS and the DDT 2756 programs reported the absence of \lya{} \citep{Morishita2023}. This was a surprising result at the time, as the strong overdensity suggested that there may be a large bubble and enhanced Ly$\alpha$ transmission. 
Utilizing deep prism spectra from the UNCOVER observations,  \cite{Chen2024} detected \lya{} emission in one of the newly-confirmed member galaxies (UNCOVER-23604 at $z=7.88$) and strong damped \lya{} absorption in three  galaxies. 
UNCOVER-23604 is  the only \lya{} emitter among the 19 galaxies in our sample (excluding the $z=7.03$ AGN; \citealt{Furtak2023}). 
Because this is a compact grouping of galaxies, the typical separation between members is small ($\lesssim60$~pkpc). As a result, \citet{Chen2024} suggest that the absence of Ly$\alpha$ emission (and the presence of damped Ly$\alpha$ absorption) may be driven by 
neutral hydrogen in member galaxies and tidally-disrupted material from interactions. Efforts to link Ly$\alpha$ emission to ionized bubble sizes are better suited to overdensities spanning larger physical  scales ($>$ 1 pMpc, similar to those described in previous subsections), where member galaxy separations are greater than those of the Abell 2744 structure.

\section{Discussion}\label{sec:results_lya}

\tabletypesize{\scriptsize}
\begin{deluxetable*}{lccccccccccc}
\tablecaption{Summary of identified large-scale ($\sim$~pMpc) spectroscopic overdensities and overdensity candidates across the fields. For each structure, we list whether it is identified from NIRSpec or NIRCam grism observations (Type), the redshift range, the number of NIRSpec galaxies, the number of \lya{} emitters, the number of high-EW \lya{} emitters (EW$>$100~\AA{}), the redshift span (d$z$), the corresponding line of sight distance ($d_{\rm LOS}$), the projected area in arcmin$^2$ and pMpc$^2$, as well as the estimate of the overdensity factor.}\label{tab:overdensity_summary}
\tablehead{
\colhead{Structure}           & \colhead{Type}   & \colhead{$z$}        & \colhead{N sources} & \colhead{N LAEs} & \colhead{N high-EW} & \colhead{d$z$} & \colhead{$d_{\rm LOS}$} & \colhead{Area}           & \colhead{Area}           & \colhead{$N/\langle N \rangle$} \\
\colhead{}                   & \colhead{}       & \colhead{}           & \colhead{}          & \colhead{}       & \colhead{}           & \colhead{}            & \colhead{(pMpc)}         & \colhead{(arcmin$^2$)}   & \colhead{(pMpc$^2$)}     & \colhead{}
}
\startdata
UDS $z\simeq7.1$             & NIRSpec          & 7.08--7.16           & 11 &  0           &  0           & 0.08 & 2.1 & $10.3\times2.9$   & $3.2\times0.9$       & $>1.6$                 \\
UDS $z\simeq7.4$             & NIRSpec          & 7.24--7.43           & 22 &  2           &  1           & 0.19 & 7.1 & $11.3\times5.9$   & $3.5\times1.8$       & $>2.2\mbox{--}4.7$\tablenotemark{a} \\
UDS $z\simeq7.8$             & NIRSpec          & 7.74--7.83           & 11 &  1           &  1           & 0.09 & 3.2 & $8.3\times2.7$    & $2.4\times0.8$       & $>4.2$                        \\
EGS $z\simeq7.0$\tablenotemark{b} & NIRSpec      & 7.00--7.06           & 13 &  0           &  0           & 0.06 & 2.6 & $7.0\times3.5$    & $2.2\times1.1$       & $>3.2$                        \\
EGS $z\simeq7.2$             & NIRSpec          & 7.16--7.20           & 16 &  2(3)\tablenotemark{c} & 1(2)\tablenotemark{c} & 0.04 & 1.6 & $10.3\times5.0$   & $3.2\times1.5$       & $>3.5$                        \\
EGS $z\simeq7.4$             & NIRSpec          & 7.38--7.56           & 22 &  7           &  1           & 0.18 & 6.6 & $18.6\times6.0$   & $5.6\times1.8$       & $>2.8\mbox{--}4.5$\tablenotemark{d} \\
EGS $z\simeq7.9$             & NIRSpec          & 7.90--7.99           &  6 &  0           &  0           & 0.09 & 2.9 & $8.3\times1.4$    & $2.4\times0.4$       & $>3.3$                        \\
GOODS-S $z\simeq7.2$         & Grism            & 7.16--7.36           & 19 &  3           &  1           & 0.20 & 8.0 & $R\sim4.4$        & $R\sim1.4$           & $4.0$                         \\
GOODS-S $z\simeq7.6$         & Grism            & 7.52--7.72           &  7 &  2           &  0           & 0.20 & 7.2 & $R\sim4.4$        & $R\sim1.3$           & $4.5$                         \\
GOODS-S $z\simeq7.9$         & NIRSpec          & 7.89--8.07           & 16 &  6           &  1           & 0.18 & 5.9 & $4.4\times3.8$    & $1.3\times1.1$       & $>5.1$                        \\
GOODS-S $z\simeq8.2$         & NIRSpec          & 8.20--8.28           &  6 &  4           &  0           & 0.08 & 2.3 & $4.1\times0.5$    & $1.2\times0.1$       & $>60$                         \\
GOODS-N $z\simeq7.1$         & Grism            & 7.00--7.20           & 17 &  2           &  1           & 0.20 & 8.4 & $R\sim4.4$        & $R\sim1.4$           & $11.0$                        \\
GOODS-N $z\simeq7.6$         & Grism            & 7.48--7.68           &  0 &  0           &  0           & 0.20 & 7.2 & $R\sim4.4$        & $R\sim1.3$           & $4.5$                         \\
\enddata
\tablecomments{
a. Overdensity factors are calculated for the two substructures at $z=7.24\mbox{--}7.32$ and $z=7.36\mbox{--}7.43$.\\
b. This structure likely extends down to $z=6.93$ with the rest of the sources listed in Table~\ref{tab:egs_z6p9}.\\
c. Numbers in the brackets are when we include CEERS-44 that is near the edge of the footprint in this structure.\\
d. Overdensity factors are calculated for the two substructures at $z=7.43\mbox{--}7.47$ and $z=7.45\mbox{--}7.49$.
}
\end{deluxetable*}

We have characterized the environment and Ly$\alpha$ properties of 292 
galaxies at $z=7.0-8.5$ spanning five fields and a total area of $\sim453$~arcmin$^2$ (total volume $1.3\times10^6$~cMpc$^3$), identifying 36 Ly$\alpha$ emitters and 13 likely overdense large scale structures. In this 
section, we explore constraints on the sizes of the ionized regions that may be implied by the statistical Ly$\alpha$ properties. Here we take a simple approach, but we will discuss the potential of new methods \citep{Nikolic2025,Lu2024_edge} which should yield robust bubble sizes as 
{\it JWST} spectroscopic datasets increase in size and sensitivity. 

We first consider the range of bubble sizes that are likely to be present in our survey volume based on recent theoretical work (\citealt{Lu2024}) investigating   
semi-numerical simulations of reionization \citep{Mesinger2011,Mesinger2007,Sobacchi2014}.  The typical bubble size depends on the stage of reionization (e.g., \citealt{Furlanetto2005,Mesinger2007,Geil2016,Lin2016}), with small sizes present when neutral hydrogen fractions are large, and larger sizes appearing as the IGM becomes more ionized. At fixed $\overline{x}_{\rm HI}$, the sizes depend on the mass scale of the dominant 
ionizing source population, with larger bubbles found when reionization is driven by massive sources. For standard source models (where reionization is primarily driven by galaxies in low mass halos, $M_h \gtrsim 10^8 M_\odot$, defined as ``gradual"), \cite{Lu2024} find that typical bubble sizes increase from 0.3 pMpc ($\overline{x}_{\rm HI}=0.8$) to 2.3 pMpc ($\overline{x}_{\rm HI}=0.4$; see also \citealt{Mesinger2007,Seiler2019}). 

At the redshifts we consider in this 
paper, the IGM neutral fraction is expected to be in the range $\overline{x}_{\rm HI}$=0.5--0.7 (e.g., \citealt{{Tang2024_nirspec,Kageura2025,Mason2025}}). 
Considering an IGM neutral fraction of 
$\overline{x}_{\rm HI}$=0.6 at $z\sim7$, \cite{Lu2024} predicts the bubbles surrounding galaxies with $M_{\rm UV} < -19.0$ (similar to those in our sample) to have a median radius ($R_{\rm ion}$) of 1.1~pMpc, with an inner 50\% range of 0.4--2.1~pMpc. This suggests that the bubbles at $z\simeq 7$ 
should have radial diameters of d$z=0.05$ (average) with the inner 50\% range corresponding to d$z=0.02-0.10$. The angular extent of the bubble radii
corresponds to 3.5 arcmin (average) and 1.4--6.8 arcmin for the inner 50\% range. The {\it JWST} imaging fields often span on order 10 arcmin on a side, and hence we may find large fractions of individual fields covered by single ionized bubbles at $z\simeq 7$. Given the relatively small areal coverage, {\it JWST} spectroscopy is best suited to probing the radial structure of bubbles. 
In particular, we expect to see strong overdensities and Ly$\alpha$ emitters grouped in bins of d$z=0.05$ for typical bubble sizes. If we identify overdensities and enhanced Ly$\alpha$ emission spanning larger redshift bins (i.e. d$z=0.4$ or R=6--8 pMpc), this would indicate large 
bubbles not expected in the standard (``gradual") reionization model adopted in \citet{Lu2024} and other recent studies. In the following, we investigate the likely sizes of the ionized structures and overdensities in the $\simeq 10^6$ cMpc$^3$ of volume analyzed in this paper.

\subsection{The Environment of Strong Ly$\alpha$ Emitters}

We expect the galaxies with the largest Ly$\alpha$ EWs and Ly$\alpha$ escape 
fractions to trace ionized sightlines. 
Among the 36 Ly$\alpha$ detections, 9 are found with  \lya{} emission  EWs exceeding $100$~\AA{}, close to the intrinsic value expected for young stellar populations (e.g. \citealt{Charlot1993,Chen2024}). 
Nearly all (8/9) of the strongest Ly$\alpha$ emitters are located in the overdense structures we have identified, as would be expected if the overdense regions are carving out ionized bubbles boosting the transmission of Ly$\alpha$. The overdensities we have identified at $z\simeq 7.0-8.5$ have redshift widths of d$z\simeq 0.04\mbox{--}0.2$, corresponding to line-of-sight proper distances of 2--8 pMpc.  The structures appear to be spread over at least 4--11 arcmin, again suggesting size scales of 1.3--3.5 pMpc. The inferred sizes of the overdensities  (albeit somewhat uncertain with the limitations of current datasets) are consistent with the typical bubble sizes we expect at $z\simeq 7$. 
We can investigate whether the strongest Ly$\alpha$ emitters are fractionally more common in overdense regions. We find that the strong \lya{} emitters (EW$>100$~\AA{}) account for $6.4\pm2.2$\% (8/125) of the galaxies associated with the identified overdense structures. Here we have only counted \lya{} non-detections if the upper limits reach below 100~\AA{}.
If we consider galaxies that are not part of any identified structures, we find that the fraction of strong Ly$\alpha$ emitters is only $1.4\pm1.4$\% (or 1/73). This suggests that overdense regions are more likely to host the strongest Ly$\alpha$ emitters, as would be expected if these regions tended to have larger ionized sightlines.

Ly$\alpha$ emission offers the potential to begin measuring the sizes of 
the ionized regions surrounding the overdensities.
Recent work has constrained the length of the ionized sightlines associated with the strong Ly$\alpha$ emitters by quantifying their  \lya{} escape fractions (e.g., \citealt{Chen2024,Saxena2023,Witstok2024,Witstok2025_z8,Tang2024_nirspec}).  Using the methods outlined in \S~\ref{subsec:lya_measurement}, we find that the strong Ly$\alpha$ detections have escape fractions between f$_{\rm{esc,Ly\alpha}}$=0.22 and 0.71 (median =0.44). While these measurements are subject to uncertainties \citep[e.g.,][]{Chen2024,McClymont2024,Scarlata2024,Yanagisawa2024,Tang2024_z56}, the large values suggest that the bulk of the line is transmitted without 
attenuation. If we assume Ly$\alpha$ faces negligible attenuation from the galaxy interstellar medium and circumstellar medium, then the IGM attenuation is equal to 1-f$_{\rm{esc,Ly\alpha}}$ and IGM transmission is f$_{\rm{esc,Ly\alpha}}$. 
The inferred IGM transmission can be directly related to the distance to the nearest neutral intergalactic hydrogen atoms by calculating the opacity 
provided to the Ly$\alpha$ profile that emerges from a galaxy, taking into account resonant scattering and damping wing attenuation (see \citealt{Gunn1965,Miralda-Escude1998,Mason2018,Mason2020_bubble,Tang2024_nirspec}). Following this procedure, the escape fractions we measure in the strong Ly$\alpha$ emitters indicate typical ionized sightlines of at least 1 pMpc, similar to the values reported in other papers (e.g., \citealt{Saxena2023,Witstok2025_z8,Tang2024_nirspec}) and to expectations 
for the typical bubble sizes we expect at $z\simeq 7$. 

Several caveats must be noted with the approach described above. First, 
the inferred IGM transmissions (and hence ionized sightline sizes) are lower limits, as some of the Ly$\alpha$ emission is likely attenuated in the galaxy. Second, uncertainties in recombination physics \citep[e.g.,][]{Scarlata2024} are such that 
Ly$\alpha$ escape fractions may be overestimated by factors of several, significantly impacting the computed ionized 
sightline sizes. And third, the ionized sightlines we calculate using this method should not be considered as bubble radii, as the galaxies are not 
necessarily located in the center of the bubble. Nevertheless, the measurements presented here give us an  estimate of the minimum  
ionized size scale (expected under case B recombination) that may be linked to the overdensities hosting strong 
Ly$\alpha$ emitters. 

Within a given ionized bubble, we expect the Ly$\alpha$ opacity to 
depend on the distance of the galaxy to the neutral IGM along the line of sight \citep[e.g.][]{Mesinger2015}.  As a result, we expect galaxies on the far side of the bubble (along our viewing angle) to face reduced attenuation \citep[e.g.,][]{Lu2024_edge,Nikolic2025}.
To investigate this effect, we consider the relative positions of the largest EW \lya{} ($>$100~\AA) emitters  relative to the overdensities they are associated with. 
Here, we do not consider CEERS-44, which as mentioned in \S~\ref{subsec:overden_egs}, is likely to be part of the extended overdensities in EGS but the relative position within it is less clear as it lies near the edge of the footprint.
This leaves us with 7 high-EW \lya{} detections and the overdensities they are part of have been described in \S~\ref{sec:overden}.
We find that all of the 7 strong \lya{} detections are located near the center or at the back end of overdensity structures, with the typical line-of-sight distance to the front end of 3.8~pMpc (d$z$=0.09). This is consistent with what would be expected if the volume spanned by the overdense structures is ionized, again indicating typical ionized sightline sizes matched to the overdensity scale of several pMpc.

\subsection{The Transmission of Ly$\alpha$ in Ionized Bubbles}

\begin{figure*}
    \centering
    \includegraphics[width=0.49\linewidth]{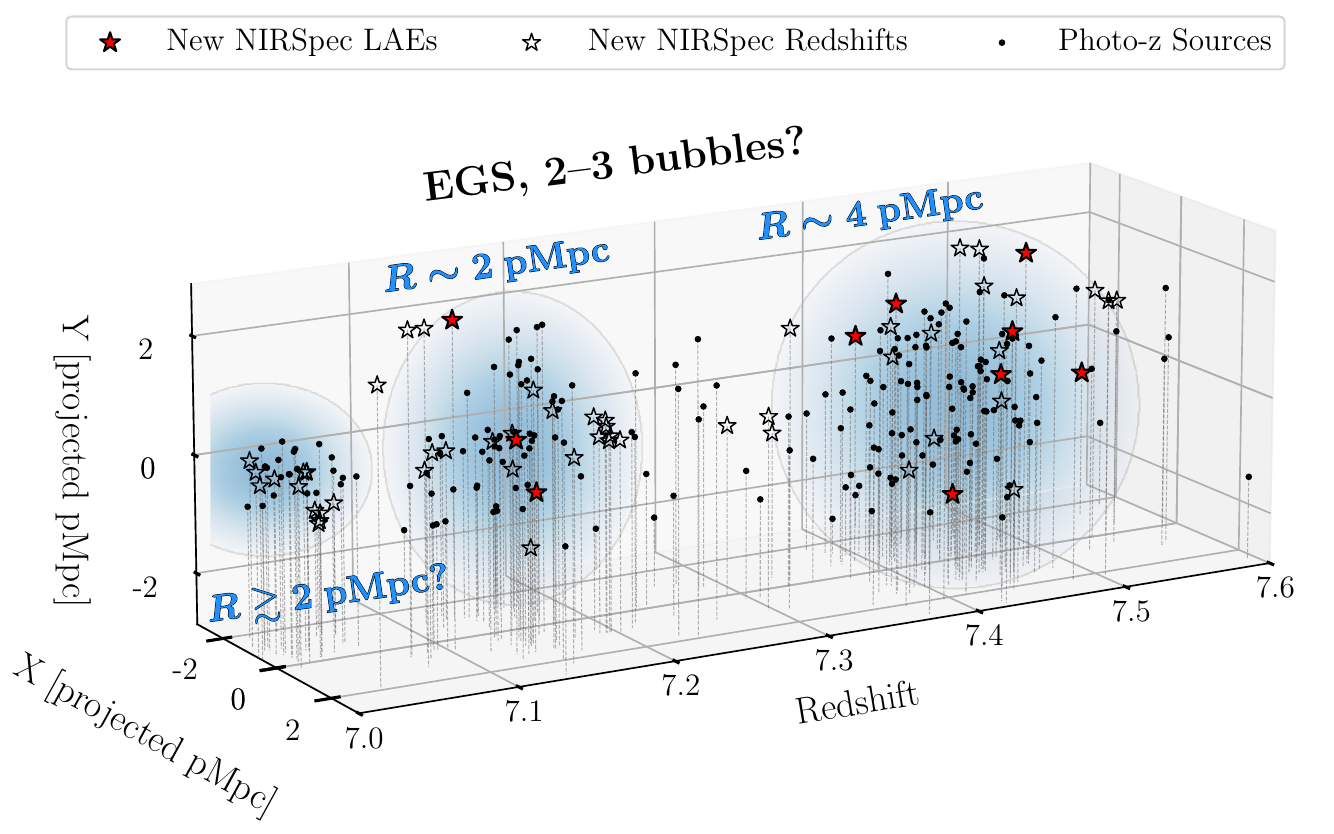}
    \includegraphics[width=0.49\linewidth]{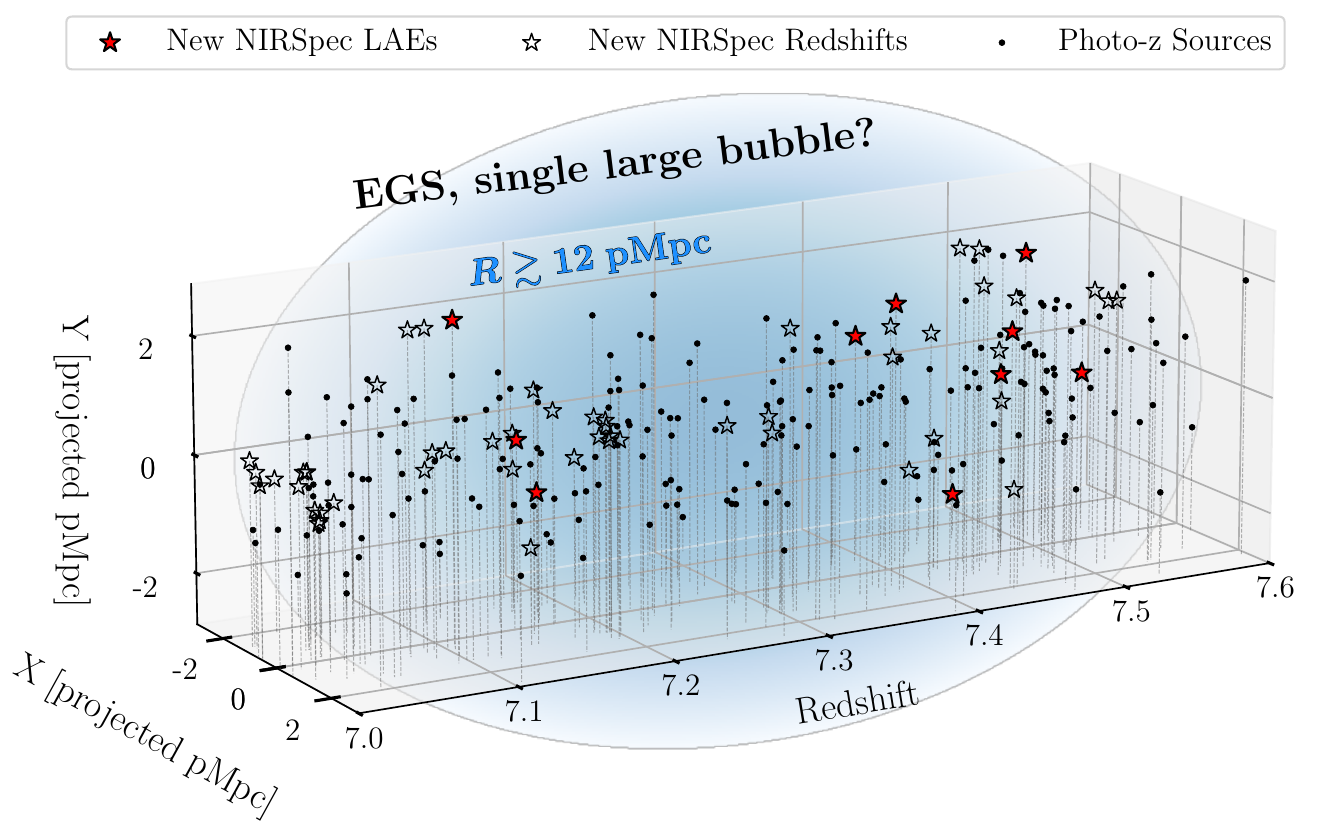}
    \caption{Illustration of likely ionized bubble morphology in the EGS field at $z=7.0\mbox{--}7.6$.
    The presence of 10 \lya{} emitters (3 at EW$>$100~\AA{}) suggests this region to be highly ionized, possibly hosting 2--3 bubbles with radii $R\sim2\mbox{--}4$~pMpc (left panel).
    Alternatively, the entire region may be a single large bubble extending over $R \gtrsim 12$~pMpc  (right panel).
    Both scenarios are extremely rarely seen in standard reionization simulations. 
    We plot the photometrically selected sources as black points, with their redshift distributed differently according to the likely position and size of the ionized bubbles.
    We also show sources already confirmed with NIRSpec as open stars and those with \lya{} detections as filled red stars.
    .}
    \label{fig:egs_bubble}
\end{figure*}

Theoretical work has begun to explore the constraining power of statistical  \lya{} spectroscopy samples on the spatial extent of the ionized structures \citep[e.g.,][]{Lu2024_edge,Nikolic2025}.
\cite{Lu2024_edge} developed methods to map the bubbles both in the plane of the sky and along the line of sight by leveraging spatial variations in \lya{} transmission inferred from the equivalent width (EW) measurements of individual galaxies.
In this framework, bubble edges in the plane of the sky are identified by sharp declines in Ly$\alpha$ transmission, while the extent along the line of sight is constrained by gradients in transmission as a function of line-of-sight distance.
\citet{Nikolic2025} introduced a complementary forward-modeling approach, in which the spatially varying IGM transmission is modeled for each assumed combination of bubble position and size.
For an ensemble of galaxies, the bubble parameters are constrained by identifying the model that best reproduces the observed \lya{} EWs given the source positions. These methods will become possible as larger spectroscopic 
samples emerge. 

As a first step with our existing statistical database, we investigate the average transmission that the IGM provides 
to Ly$\alpha$ in individual overdensities.
To do so, we compare the observed \lya{} EWs to the average \lya{} EW distribution at $z\sim5$ from \cite{Tang2024_z56}, a redshift at which the IGM damping wing attenuation is expected to be minimal.
This $z\sim5$ intrinsic EW distribution is described by a log-normal function with parameters $\mu=2.38_{-0.31}^{+0.28}$ (corresponding to a median EW of $10.8_{-2.9}^{+3.5}$~\AA{}) and $\sigma=1.63_{-0.19}^{+0.23}$ \citep[e.g.,][]{Tang2024_z56,Tang2024_nirspec}.
We infer the IGM transmission, $\mathcal{T}$, within each structure using a Bayesian framework.
While we outline the main steps below, our approach closely follows that of \citet{Tang2024_z56} previously developed to infer the Ly$\alpha$ EW distribution.
We adopt a flat prior over $0\leq\mathcal{T}\leq1$ for the average transmission within each structure.
The likelihood of a given $\mathcal{T}$ is computed as the product of the individual likelihoods for all EW measurements in the sample, where we account for both \lya{} detections and non-detections (upper limits).
For each \lya{} detection, we incorporate the measurement uncertainty by modeling the observed EW as a Gaussian-distributed variable in the likelihood calculation.
For non-detections, we compute the likelihood as the probability that the EW falls below the $3\sigma$ upper limit, assuming the intrinsic EW distribution modulated by the transmission factor $\mathcal{T}$.
We compute the average IGM transmission using Markov Chain Monte Carlo (MCMC) sampling through the \textsc{emcee} package \citep{Foreman-Mackey2013}.
From the resulting posterior distribution, we compute the median transmission and the uncertainties corresponding to the marginalized 68\% credible intervals.

To get a sense of current constraints, we consider two overdensities identified with grism observations: the $z\simeq7.2$ structure in GOODS-S and the $z\simeq7.1$ structure in GOODS-N.
Each structure hosts one large EW \lya{} emitter, along with 14--17 other galaxies with NIRSpec \lya{} constraints, which include 1--2 weaker \lya{} emitters.
As expected from the presence of strong \lya{} detections, both regions are consistent with transmitting a significant fraction of \lya{} photons: $\mathcal{T}=0.52_{-0.22}^{+0.28}$ for the $z\simeq7.2$ structure in GOODS-S, and $\mathcal{T}=0.55_{-0.22}^{+0.23}$ for the $z\simeq7.1$ structure in GOODS-N.
We translate the IGM transmission into ionized bubble sizes ($R_{\rm ion}$) by modeling the expected Ly$\alpha$ optical depth from both resonant scattering and the damping wing, following the methodology described in \citet{Tang2024_nirspec}.
While uncertainties are substantial, the estimated transmission values are consistent with expectations for large ionized bubbles (for $R_{\rm ion}=2$~pMpc, $\mathcal{T}=0.61$).
Deeper Ly$\alpha$ spectroscopy of galaxies confirmed by the NIRCam grism in these overdensities (8 in GOODS-S, 22 in GOODS-N; \citealt{Meyer2024}) and selected photometrically ($\sim120$ in GOODS-S, $\sim60$ in GOODS-N; \citealt{Endsley2024_jades})  could expand the current sample in these regions by a factor of 4–-6, enabling more robust constraints on the extent of ionized bubbles.

\subsection{A Large Ionized Region in the EGS?}\label{subsec:egs_ionized}

\begin{figure*}
    \centering
    \includegraphics[width=1\linewidth]{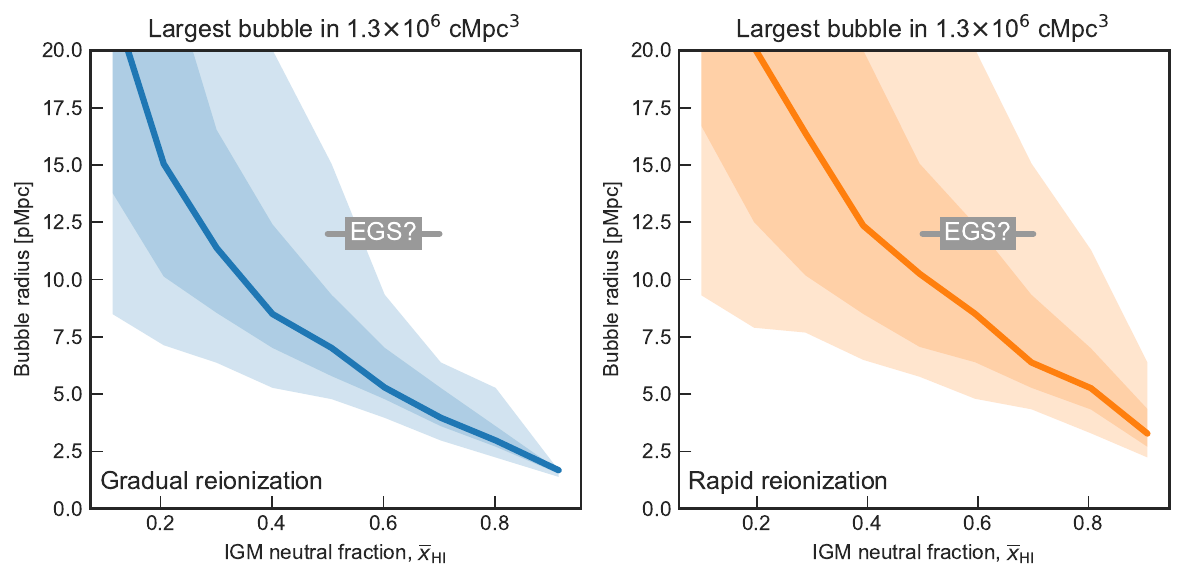}
    \caption{Radius of the largest ionized bubble expected within our survey volume ($1.3\times10^6$~cMpc$^3$) as a function of mean IGM neutral fraction ($\overline{x}_{\rm HI}$).
    We show predictions from simulations by \citet{Lu2024}, considering both the fiducial `gradual' reionization model (left) and the `rapid' model (with reionization driven by galaxies in rarer, more massive halos, right panel).
    We estimate the radius of the largest bubble from 1000 realizations and plot their median values in thick lines, with the shallower and darker shaded regions corresponding to the inner 68\% and 95\% percentiles.}
    \label{fig:largest_bubble}
\end{figure*}

In the previous subsections, we have demonstrated that current data are consistent with overdensities and ionized bubbles spanning several pMpc in size across a volume of 1.3$\times10^6$ cMpc$^3$ at $z\simeq 7.0-8.5$. These measurements are similar to predictions for the average bubble sizes expected near the midpoint of reionization \citep{Lu2024}. It is possible that our survey volume intersects one or more bubbles that are much larger than this value. Identification of any bubbles with radii in excess of 8 pMpc may provide hints that the source population differs from that often assumed in reionization models. With our current dataset, we should be able to pick such structures out by their radial  (d$z$=0.4--0.5 in diameter) and 
tangential ($\approx25$ arcmin in diameter) sizes. 

Attention has long focused on the Ly$\alpha$ emitters in the EGS field (e.g., \citealt{Oesch2015,Roberts-Borsani2016,Stark2017,Tilvi2020,Jung2022,Tang2023,Chen2024,Napolitano2024}). {\it JWST} has already demonstrated that Ly$\alpha$ transmission is enhanced in this field (\citealt{Napolitano2024,Tang2024_nirspec}). With the larger spectroscopic sample presented in this paper, we are now able to obtain an improved 
view of the redshift distribution of galaxies. We have identified three 
overdense structures between $z\simeq 7.0$ and $z\simeq 7.6$ (a radial distance of 24 pMpc), and we note that there is additional evidence 
for a significant structure (28 additional galaxies) extending to $z\simeq 6.9$ (Appendix~\ref{appedix:egs_z6p9}). 
Moreover, between $z\simeq 7.1$ and $z\simeq 7.5$, we find three extremely strong (EW$>$100~\AA) Ly$\alpha$ emitters. Over 
$z\simeq 7.10-7.56$, we find 10 Ly$\alpha$ emitting galaxies. 
This means that 28\% of the Ly$\alpha$ emitters in our sample lie in a region of the EGS spanning 18 pMpc  in length. 

If there is a very large bubble in the five fields, it is likely in the EGS between $z\simeq 7.0$ and $z\simeq 7.6$ (Figure~\ref{fig:egs_bubble}). It is possible we are seeing several (2--3) bubbles along the line-of-sight, perhaps corresponding to an extended filamentary structure, with multiple overdense structures spanning $\sim24$ pMpc. It is also possible we are seeing one extremely large bubble with a radius of $\gtrsim$12 pMpc. With current data, we cannot distinguish between these scenarios. But the combination of wide-field grism observations and additional Ly$\alpha$ spectroscopy should make it clear 
which picture is correct. 

Should we be surprised to find such a large bubble or filament in our survey area at $z\simeq 7$?
We test the likelihood of finding large ionized regions by comparing with predictions from the fiducial `gradual' reionization simulations and `rapid' simulations (where reionization is driven by galaxies in rarer, more massive halos) by \citet{Lu2024}. 
We estimate the expected size distribution of bubbles in our survey volume, and within the EGS field, by sampling from the bubble size distributions derived by \citet{Lu2024} in Gpc-scale IGM simulations. 
We calculate the volume of our survey which is expected to be ionized at a given IGM neutral fraction $\overline{x}_{\rm HI}$, $V_{\rm ion} =(1-\overline{x}_{\rm HI})\times V$, and sample bubbles from the corresponding bubble size distributions until the ionized volume is filled.
To account for large bubbles extending outside the survey volume we allow the final bubble sampled to `overfill' the volume, but only count it if the volume of the bubble inside the survey volume corresponds to at least d$z=0.2 \times 100$\,arcmin$^2$. This is comparable to the minimum volume of overdensities in EGS and so allows us to include large bubbles which may impact Ly$\alpha$ visibility, but avoids counting large bubbles which only intersect a tiny fraction of the survey volume. We note our conclusions are unchanged if we consider smaller overlap volumes, and if we sample volumes directly from the simulated IGM cubes. 
We repeat for 1000 iterations to find the distribution of bubble sizes expected in our survey volume and in the volume of EGS potentially containing large bubbles (see below). 

However, in the smaller volume of EGS where we identify potentially 3 overdensities hosting LAEs within $z=7.0-7.6$ over $\approx 129\,$arcmin$^2$ of NIRSpec coverage (corresponding to $1.7\times10^5$\,cMpc$^3$), we would expect just $0-1$ $R\gtrsim$2\,pMpc bubbles in either reionization scenario. Finding multiple large bubbles in such a small volume is rare: we find $<20\%$ of simulated EGS volumes host 2 $R\gtrsim$2\,pMpc bubbles and $<1\%$ of volumes host 3 such large bubbles.

Based on these models, it may be more likely that the $z\approx7.0-7.6$ line-of-sight in EGS is spanned by a single, large ($R \gtrsim 12$\,pMpc) ionized region. We calculate the largest bubble expected to be overlapping with our full survey volume, $V=1.3\times10^6$\,cMpc$^3$, from our 1000 realizations of the bubble size distribution above (Figure~\ref{fig:largest_bubble}).
We find it is extremely unlikely to find a $R \gtrsim 12$\,pMpc bubble within our survey volume in the `gradual' reionization scenario, assuming the IGM is $\gtrsim50$\% neutral. We find $<5$\% of sampled volumes should overlap with such large bubbles assuming $\overline{x}_{\rm HI}\approx0.5-0.7$, and we find \textit{no} bubbles that large in our sampled volumes when $\overline{x}_{\rm HI}\gtrsim0.7$. 
By contrast, while the largest bubbles are typically $6-10$\,pMpc (inner 50\% range) in the `rapid' reionization scenario, 25\% of sampled volumes in this model contained bubbles with $R \gtrsim 12$\,pMpc.
Thus, while the number of observed overdensities around LAEs in our survey volume appears more consistent with the standard `gradual' reionization model, the existence of either multiple $R\gtrsim2$ pMpc bubbles or a large $R \gtrsim 12$\,pMpc ionized region in EGS is extremely challenging to explain in that scenario.
Establishing the extent of the ionized region in EGS with more complete spectroscopy over a wider area would have important implications for our understanding of reionization.

\section{Conclusions}

In recent years, \jwst{} spectroscopy has rapidly advanced the study of \lya{} emission in the reionization era.
The expanding dataset now allows for the characterization of the large scale environment of \lya{} emitters, providing insight into the emergence of large ionized structures amid a significantly neutral IGM.
In this work, we investigate the spatial distribution of \lya{} emitters relative to galaxy overdensities at $z=7.0\mbox{--}8.5$ over five independent fields.
Our main findings are summarized below.

(i) From our uniform reduction of the publicly-available \jwst{}/NIRSpec observations, we assemble a spectroscopic sample of 292 galaxies at $z=7.0\mbox{--}8.5$ across five independent fields: UDS, EGS, GOODS-S, GOODS-N, and Abell 2744.
A significant fraction of these observations were taken in Cycles 2 and 3, yielding a final sample size that is $>3\times$ larger compared to earlier analysis at these redshifts \citep[e.g.,][]{Napolitano2024,Tang2024_nirspec,Kageura2025}.
    
(ii) We detect \lya{} emission in 36 galaxies at $z=7.0\mbox{--}8.5$ from our sample, many of which have also been reported in previous studies \citep[e.g.,][]{Napolitano2024,Tang2024_nirspec,Kageura2025,Jones2025}.
Nine of the \lya{} detections are newly presented here, all of which are from spectra obtained during Cycles 2 and 3.
Notably, two galaxies in the UDS field show very strong \lya{}: RUBIES-UDS-24303 (EW = 200~\AA{}) and CAPERS-UDS-142615 (EW = 284~\AA{}), adding to the still small but growing sample of high-EW \lya{} emitters identified at $z>7$ \citep[e.g.,][]{Nakane2023,Napolitano2024,Saxena2023,Chen2024,Tang2024_nirspec,Witstok2025_z8}. 

(iii) We utilize spectroscopic redshifts from NIRCam grism and NIRSpec observations to characterize the large scale environments across the five fields.
Within a comoving volume of total volume $1.3\times10^6$~Mpc$^3$, we find in total 13 overdensities or likely overdense regions spanning d$z$=0.04--0.20, corresponding to line of sight distances of 2--8~pMpc.
These structures extend over angular scales of at least 4--11~arcmin, implying projected physical distances of $\gtrsim$1.2--3.5 pMpc.
They are found to be significantly ($\gtrsim$2 up to $60\times$) overdense relative to predictions from the UV luminosity function or the average number densities of the corresponding grism surveys.
We additionally find evidence for photometric overdensities that overlap with the structures in UDS and EGS, further supporting the presence of large scale overdense regions in these fields.

(iv) Our sample includes nine strong \lya{} emitters with rest-frame EWs exceeding 100~\AA{}, the majority of which (8/9) are located within one of the structures we found.
The uniform association between strong \lya{} detections and galaxy overdensities is consistent with the interpretation that these regions can carve out large ionized sightlines, enabling efficient transmission of \lya{} photons through the IGM. 
We further find that these strong \lya{} emitters are almost exclusively situated near the center or the back end of the corresponding structures, with the typical distance of 3.8~pMpc to the front end, which is consistent with expectations if the entire region is ionized \citep{Lu2024_edge,Nikolic2025}.

(v) The overdensities with high-EW \lya{} detections span physical scales comparable to the ionized bubble sizes predicted by reionization simulations (inner 50\% range of $R_{\rm ion}$=0.4--2.1 pMpc, assuming $\overline{x}_{\rm HI}=0.6$).
We consider the current constraints of the IGM transmission within the two grism overdensities, and the derived values are consistent with the presence of large ionized bubbles ($R_{\rm ion}\sim2$~pMpc).
Deep \lya{} spectroscopy of  more galaxies (photometrically-selected and grism-confirmed) within this field will be crucial to narrow down the likely range of bubble sizes.

(vi) We show that the EGS field is likely highly ionized over the  redshift range of 7.0--7.6 given the presence of three galaxy overdensities in addition to 10 \lya{} emitters, three of which have EWs exceeding 100~\AA{}.
It is possible we are seeing evidence for 2--3 $R_{\rm ion}=$2--4~pMpc bubbles along the line of sight or a single large ionized bubble with $R_{\rm ion}\gtrsim12$~pMpc.
Both cases are in tension with  standard reionization models.
Wide area redshift surveys with more complete \lya{} spectroscopy will help map the physical extent of ionized structures in this field.

\begin{acknowledgments}

The authors thank Stéphane Charlot and Jacopo Chevallard for providing access to the \beagle{} tool used for SED fitting analysis.
DPS acknowledges support from the National Science Foundation through the grant AST-2109066.
CM acknowledges support by the VILLUM FONDEN under grant 37459 and the Carlsberg Foundation under grant CF22-1322. The Cosmic Dawn Center (DAWN) is funded by the Danish National Research Foundation under grant DNRF140.
M. Tang acknowledges funding from the JWST Arizona/Steward Postdoc in Early galaxies and Reionization (JASPER) Scholar contract at the University of Arizona.

This work is based in part on observations made with the NASA/ESA/CSA JWST. The data were obtained from the Mikulski Archive for Space Telescopes at the Space Telescope Science Institute, which is operated by the Association of Universities for Research in Astronomy, Inc., under NASA contract NAS 5-03127 for JWST. 
These observations are associated with program GO 4287 (PI: C. Mason \& D. Stark), and the following publicly available programs GTO 1180, 1181, 1210, 1286, 1287,
and GO 3215 (JADES; doi: \href{https://archive.stsci.edu/doi/resolve/resolve.html?doi=10.17909/8tdj-8n28}{10.17909/8tdj-8n28}), ERS
1324 (GLASS; doi: \href{https://archive.stsci.edu/doi/resolve/resolve.html?doi=10.17909/kw3c-n857}{10.17909/kw3c-n857}), ERS 1345
and DDT 2750 (CEERS; doi: \href{https://archive.stsci.edu/doi/resolve/resolve.html?doi=10.17909/z7p0-8481}{10.17909/z7p0-8481}), GO 2561 (UNCOVER), DDT 2756, GO 3073, GO 4233 (RUBIES), as well as GO 6368 (CAPERS). The authors acknowledge the JADES, GLASS, CEERS, UNCOVER, DDT 2756, GO 3073, RUBIES, and CAPERS
teams led by D. Eisenstein \& N. Lützgendorf,
K. Isaak, T. Treu, S. Finkelstein, P.
Arrabal Haro, I. Labb\'e \& R. Bezanson, W. Chen, M. Castellano, A. de Graaff \& G. Brammer, and M. Dickinson
for developing their observing programs.
This research
is also based in part on observations made with the
NASA/ESA Hubble Space Telescope obtained from the
Space Telescope Science Institute, which is operated by
the Association of Universities for Research in Astronomy, Inc., under NASA contract NAS 5–26555. 
Part of the data products presented herein were retrieved from
the Dawn JWST Archive (DJA).
DJA is an initiative of the Cosmic Dawn Center, which is funded by the Danish
National Research Foundation under grant DNRF140.
This material is based in part upon High Performance Computing
(HPC) resources supported by the University of Arizona TRIF, UITS,
and Research, Innovation, and Impact (RII) and maintained by the
UArizona Research Technologies department.

This work made use of {\sc Astropy}:\footnote{\url{http://www.astropy.org}} a community-developed core Python package and an ecosystem of tools and resources for astronomy \citep{astropy:2013, astropy:2018, astropy:2022}; \beagle{} \citep{Chevallard2016}; {\sc Emcee} \citep{Foreman-Mackey2013}; {\sc Jupyter} \citep{Kluyver2016}; {\sc Matplotlib} \citep{Hunter:2007}; {\sc Numpy} \citep{harris2020array}; {\sc Photutils}, an {\sc Astropy} package for detection and photometry of astronomical sources \citep{Bradley2022}; {\sc Scikit-image} \citep{vanderWalt2014}; {\sc Scipy} \citep{2020SciPy-NMeth}; {\sc Sedpy} \citep{Johnson2021_sedpy}; and {\sc Shapely} \citep{Gillies_Shapely_2025}.

\end{acknowledgments}

%






\appendix

\restartappendixnumbering

\section{Sample Table}
\startlongtable

\begin{deluxetable}{ccccccccc} 

\tabletypesize{\scriptsize} 

\tablecaption{The sample of 292 galaxies at $z=7.0\mbox{--}8.5$ analyzed in this work.
We list the source IDs, coordinates, spectroscopic redshifts ($z_{\rm spec}$), F150W magnitudes, absolute UV magnitudes, \oiiihb{} EWs, \lya{} EWs measured from the grating and prism spectra.
We group sources by the grism overdensity or NIRSpec overdensity candidates we identified.} 

\tablehead{ 
\colhead{ID} & \colhead{RA} & \colhead{Dec} & \colhead{$z_{\rm spec}$} & \colhead{F150W} & \colhead{MUV} & \colhead{EW \oiiihb{}} & \colhead{EW \lya{}} & \colhead{EW \lya{}} \\
\colhead{} & \colhead{(deg)} & \colhead{(deg)} & \colhead{} & \colhead{(mag)} & \colhead{(mag)} & \colhead{(\AA{})} & \colhead{(Grating, \AA{})} & \colhead{(Prism, \AA{})} \\
} 

\startdata 
\multicolumn{9}{l}{\textbf{UDS: $\bm{ z\simeq 7.1 }$ Overdensity Candidate}} \\
\hline 
RUBIES-UDS-19112   &  34.352530 &  -5.288720 & 7.084 & $26.14_{ -0.04 }^{ +0.04 }$ & $-20.82_{ -0.04 }^{ +0.04 }$ & $ 874_{ - 142 }^{ + 151 }$ & --                      & --                      \\
RUBIES-UDS-132422  &  34.280574 &  -5.159774 & 7.090 & $28.34_{ -0.18 }^{ +0.22 }$ & $-18.40_{ -0.19 }^{ +0.27 }$ & $3638_{ -1197 }^{ +1572 }$ & --                      & $<196$                  \\
RUBIES-UDS-51816   &  34.274045 &  -5.224737 & 7.091 & $27.84_{ -0.15 }^{ +0.17 }$ & $-19.11_{ -0.15 }^{ +0.17 }$ & $1224_{ - 221 }^{ + 325 }$ & --                      & --                      \\
RUBIES-UDS-959606  &  34.265806 &  -5.224323 & 7.097 & $26.73_{ -0.07 }^{ +0.07 }$ & $-20.21_{ -0.07 }^{ +0.08 }$ & $ 568_{ - 145 }^{ + 123 }$ & --                      & $< 54$                  \\
RUBIES-UDS-951669  &  34.281611 &  -5.237581 & 7.105 & $26.43_{ -0.05 }^{ +0.06 }$ & $-20.43_{ -0.07 }^{ +0.07 }$ & $2045_{ - 289 }^{ + 327 }$ & --                      & $< 72$                  \\
RUBIES-UDS-138196  &  34.247828 &  -5.151997 & 7.110 & $27.54_{ -0.10 }^{ +0.11 }$ & $-19.20_{ -0.10 }^{ +0.11 }$ & $1052_{ - 130 }^{ + 153 }$ & --                      & $<181$                  \\
RUBIES-UDS-33451   &  34.352461 &  -5.264610 & 7.110 & $28.12_{ -0.16 }^{ +0.19 }$ & $-18.45_{ -0.21 }^{ +0.26 }$ & $5012_{ -1346 }^{ +1083 }$ & --                      & $<117$                  \\
\enddata 

\label{tab:sample} 

\end{deluxetable}

\section{Additional galaxies associated with the EGS $z\simeq7.0$ overdensity candidate}\label{appedix:egs_z6p9}

\begin{table}[]
    \centering
    
    \begin{tabular}{lccc}
\hline
ID & $z_{\rm spec}$ & RA & Dec \\
   &                & (deg) & (deg) \\
\hline
RUBIES-EGS-24671     & 6.930 & 215.081725 & 52.954143 \\
RUBIES-EGS-905583    & 6.931 & 215.132742 & 52.953241 \\
RUBIES-EGS-926480    & 6.932 & 215.074618 & 52.944726 \\
CEERS-1345-717       & 6.932 & 215.081437 & 52.972181 \\
CEERS-1143           & 6.933 & 215.077006 & 52.969504 \\
RUBIES-EGS-29509     & 6.938 & 214.826920 & 52.780446 \\
CEERS-1142           & 6.938 & 215.060716 & 52.958708 \\
CEERS-481            & 6.939 & 214.827785 & 52.850615 \\
RUBIES-EGS-930139    & 6.946 & 215.119655 & 52.982830 \\
RUBIES-EGS-920424    & 6.952 & 214.853004 & 52.776489 \\
RUBIES-EGS-925921    & 6.953 & 215.075054 & 52.943828 \\
RUBIES-EGS-46724     & 6.957 & 214.907754 & 52.882491 \\
RUBIES-EGS-945424    & 6.958 & 214.878422 & 52.838464 \\
RUBIES-EGS-922409    & 6.960 & 214.850856 & 52.776674 \\
CEERS-716            & 6.960 & 215.080349 & 52.993241 \\
CAPERS-EGS-67844     & 6.960 & 214.866877 & 52.918630 \\
RUBIES-EGS-915796    & 6.960 & 215.025472 & 52.892870 \\
CAPERS-EGS-86965     & 6.972 & 214.991494 & 52.960447 \\
DDT2750-445          & 6.978 & 214.941611 & 52.929130 \\
CAPERS-EGS-11333     & 6.979 & 214.942458 & 52.914723 \\
RUBIES-EGS-55604     & 6.982 & 214.983026 & 52.956001 \\
CAPERS-EGS-83923     & 6.983 & 214.985912 & 52.960388 \\
CAPERS-EGS-61688     & 6.983 & 214.889174 & 52.839679 \\
CAPERS-EGS-83869     & 6.984 & 214.985402 & 52.960381 \\
RUBIES-EGS-966103    & 6.986 & 214.985495 & 52.958748 \\
RUBIES-EGS-975193    & 6.990 & 214.967248 & 52.963630 \\
CEERS-1102           & 6.991 & 215.091047 & 52.954285 \\
RUBIES-EGS-947982    & 6.992 & 214.806511 & 52.792767 \\
CEERS-80244          & 6.999 & 214.902155 & 52.869758 \\
\hline
\end{tabular}
    \caption{Galaxies confirmed with NIRSpec at $6.93<z<7.00$ in the EGS field that are likely also associated with the $z\simeq7.0$ overdensity candidate.}
    \label{tab:egs_z6p9}
\end{table}


\bibliography{main}{}
\bibliographystyle{aasjournal}



\end{document}